\documentclass[journal]{IEEEtran} 
\pdfoutput=1 

\def\LocalIncludesDir{Includes}

\def\FigsDir{\LocalIncludesDir/Figs}

\def\BibsDir{\LocalIncludesDir/Bibs}
\def\StylesDir{\LocalIncludesDir/Styles}
\usepackage{\StylesDir/floatflt}
\usepackage{\StylesDir/fancyheadings}

\usepackage{wrapfig}
\usepackage{amsbsy,amsthm,amsfonts,amsmath,amssymb}
\usepackage[table,dvipsnames]{xcolor}
\usepackage{mdframed}
\usepackage{framed}
\usepackage{url}
\usepackage{multirow} 
\usepackage{bbm}
\usepackage{cite}
\usepackage{soul}
\usepackage{graphicx}
\usepackage{epstopdf}

\usepackage{subcaption}

\usepackage[normalem]{ulem} 
\usepackage[export]{adjustbox} 
\usepackage{array}

\def\First{1$^\mathrm{st}\,$}

\def\quadOne{\quad}
\def\quadTwo{\quad\quad}

\def\qquadFour{\qquad\qquad\qquad\qquad}

\newcommand{\myemph}[1]{\textbf{#1}}

\def\myle{\leqslant}
\def\myge{\geqslant}
\def\Eqdef{\overset {\mathrm{def}} =}

\def\Hole{\, \cdot \,}
\def\st{\; | \;}

\def\Expv{\mathsf{E}}

\def\Prob{\mathsf{Prob}}

\newcommand{\Paren}[1]{\left( #1 \right)}
\newcommand{\Brack}[1]{\left[ #1 \right]}
\newcommand{\Curly}[1]{\left\{ #1 \right\}}
\newcommand{\Angle}[1]{\left\langle #1 \right\rangle}
\newcommand{\Abs}[1]{\left| #1 \right|}

\newcommand{\Exp}[1]{e^{#1}}

\newcommand{\Op}[1]{\mathrm{\mathbf{#1}}}

\def\Trace{\Op{tr}}

\newcommand{\mysec}[1]{Section \ref{#1}}

\newcommand{\mybegeq}[1] {\begin{equation} \label{#1}}
\def\myendeq{\end{equation}}
\newcommand{\myeq}[1]{(\ref{#1})}

\newcommand{\myfig}[1]{Fig. \ref{#1}}
\newcommand{\mytable}[1]{Table \ref{#1}}

\newtheorem{theorem}{Theorem}
\newtheorem{lemma}{Lemma}
\newtheorem{corollary}{Corollary}
\theoremstyle{definition}
\newtheorem{definition}{Definition}

\theoremstyle{remark}
\newtheorem{remark}{Remark}

\theoremstyle{remark}
\newtheorem{example}{Example}

\newcommand{\mythm}[1]{Thm. \ref{#1}}
\newcommand{\mybegthm}[1]{\begin{theorem} \label{#1}}
\def\myendthm{\end{theorem}}

\newcommand{\mybeglem}[1]{\begin{lemma} \label{#1}}
\def\myendlem{\end{lemma}}

\newcommand{\mybegcor}[1]{\begin{corollary} \label{#1}}
\def\myendcor{\end{corollary}}
\newcommand{\mydef}[1]{Def. \ref{#1}}
\newcommand{\mybegdef}[1]{\begin{definition} \label{#1}}
\def\myenddef{\end{definition}}
\newcommand{\mybegrem}[1]{\begin{remark} \label{#1}}
\def\myendrem{\end{remark}}

\newcommand{\mybegex}[1]{\begin{example} \label{#1}}
\def\myendex{\end{example}}

\def\Kion{{\textrm{K}^+}}
\def\Naion{{\textrm{Na}^+}}

\colorlet{shadecolor}{green!10}

\renewcommand{\headrulewidth}{0pt}

\renewcommand{\refname}{\vspace{-0.25in}}

\begin{document}

\title{A New Statistical Model of Electroencephalogram Noise Spectra for Real-time Brain-Computer Interfaces}

%
%
%

\author{Alan~Paris$^\ast$, \emph{Student Member, IEEE}, George~Atia, \emph{Member, IEEE}, Azadeh~Vosoughi, \emph{Member, IEEE} and Stephen~A.~Berman
\thanks{Manuscript received January 27, 2016. This material is based upon work supported by the National Science Foundation under Grant CCF-1525990. \emph{Asterisk indicates corresponding author.}}%
\thanks{$^\ast$Alan Paris, M.S. is with the NeuroLogic Lab, Institute for Simulation and Training, University of Central Florida, Orlando, FL 32816 (e-mail: atparis@knights.ucf.edu) and was supported in part by the Modeling and Simulation Program, University of Central Florida and the NSF Grant CCF-1525990. George Atia, Ph.D. and Azadeh Vosoughi, Ph.D. are with the Dept. of Electrical and Computer Engineering, University of Central Florida, Orlando, FL 32816 (e-mail: \{george.atia,azadeh\}@ucf.edu). Stephen A. Berman, M.D., Ph.D. is with the College of Medicine, University of Central Florida, Orlando, FL 32827 (e-mail: stephen.berman@ucf.edu). }%
} 

\maketitle

\begin{abstract}
{\it\textbf{Objective:}}  A characteristic of neurological signal processing is high levels of noise from sub-cellular ion channels up to whole-brain processes. In this paper, we propose a new model of electroencephalogram (EEG) background periodograms, based on a family of functions which we call generalized van der Ziel--McWhorter (GVZM) power spectral densities (PSDs). To the best of our knowledge, the GVZM PSD function is the only EEG noise model which has relatively few parameters, matches recorded EEG PSD's with high accuracy from 0 Hz to over 30 Hz, and has approximately $1/f^\theta$ behavior in the mid-frequencies without infinities.
{\it\textbf{Methods:}} We validate this model using three approaches. First, we show how GVZM PSDs can arise in population of ion channels in maximum entropy equilibrium. Second, we present a class of mixed autoregressive models, which simulate brain background noise and whose periodograms are asymptotic to the GVZM PSD.  Third, we present two real-time estimation algorithms for steady-state visual evoked potential (SSVEP) frequencies, and analyze their performance statistically.
{\it\textbf{Results:}}  In pairwise comparisons, the GVZM-based algorithms showed statistically significant accuracy improvement over two well-known and widely-used SSVEP estimators.
{\it\textbf{Conclusion:}} The GVZM noise model can be a useful and reliable technique for EEG signal processing.
{\it\textbf{Significance:}} Understanding EEG noise is essential for EEG-based neurology and applications such as real-time brain-computer interfaces (BCIs), which must make accurate control decisions from very short data epochs. The GVZM  approach represents a successful new paradigm for understanding and managing this neurological noise.

\end{abstract}


\setcounter{section}{0}


\begin{figure} [h!]
\centering
\vspace{-.02\textwidth}
\includegraphics[width=0.45\textwidth]{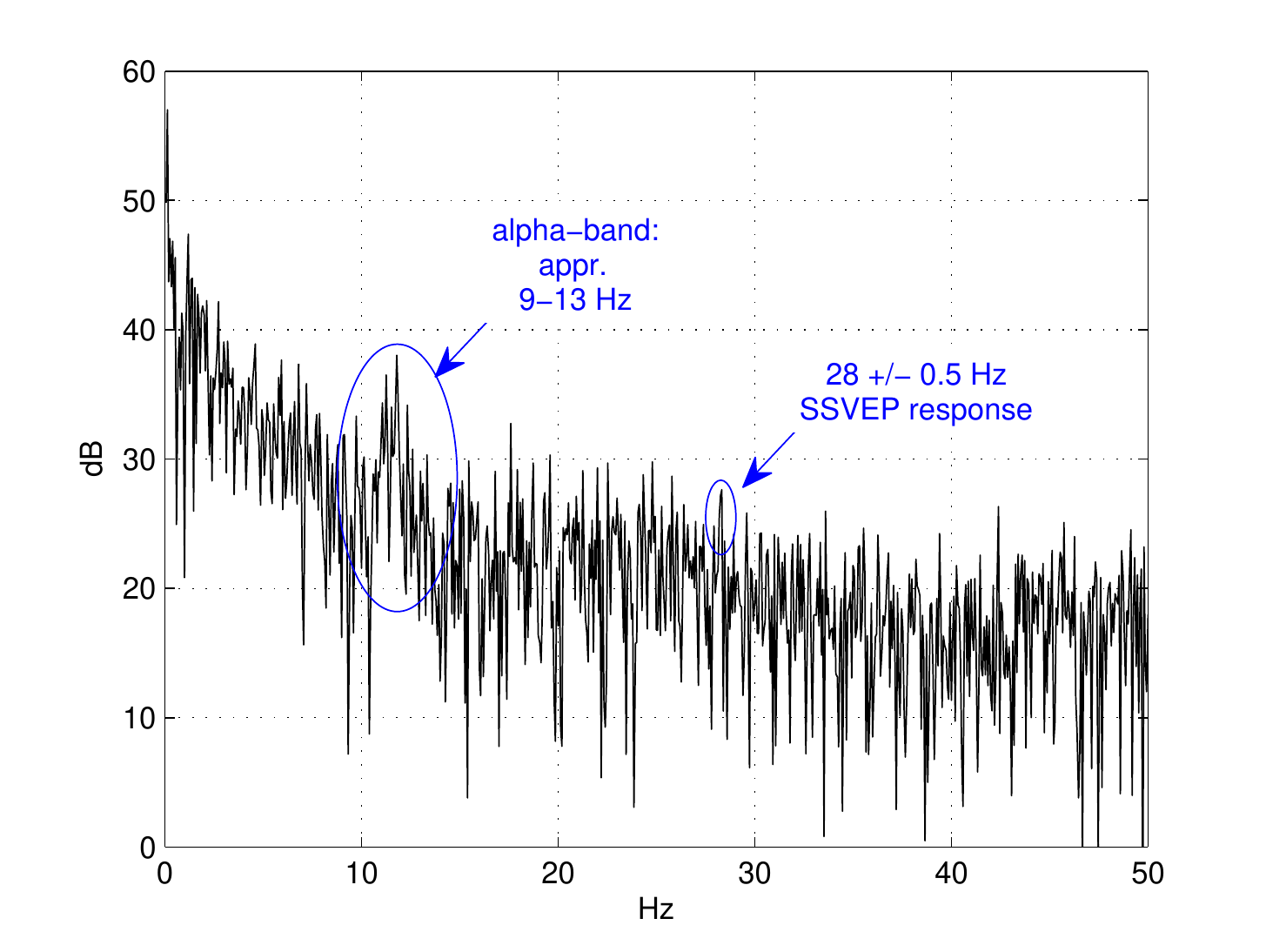}
\vspace{-.02\textwidth}
\caption{Recorded EEG periodogram from a 15-second, SSVEP experiment showing alpha-band power. The target 28 Hz response peak is nearly lost in the low SNR background noise.}
\label{fig:RawEEG28Hz}
\vspace{-.04\textwidth}
\end{figure}

\section{Introduction}\label{sec:Overview}

\subsection{Neurological Noise Research: Background and Motivation}\label{subsec:Introduction}

\IEEEPARstart{S}{ignals} recorded from living neurological tissue are extremely noisy at all scales from individual ion channels \cite{hill-chen-IV-1972} through collections of one or more neurons \cite{manwani-koch-I-1999,hille-book-2001,diba-lester-koch-2004} up to scalp-recorded EEGs \cite{nunez-srin-book-2006} (\myfig{fig:RawEEG28Hz}). As a result, the theory of neurological noise continues to be a thriving area for research \cite{dest-book-2012}, with  theoretical and practical implications for neurological signal processing and neurology.

	In this paper we examine the statistical characteristics of EEG periodograms \cite{oppen-schafer-book-1975,brockwell-davis-book-1991}. Specifically, we present a new model for the statistical properties of EEG background processes which, for the purpose of many applications, may be regarded as ``brain noise''.  To the best of our knowledge, it is the first simple and general noise model in the EEG literature which closely matches recorded EEG periodograms from near 0 Hz to over 30 Hz. We then validate this new model using three different and complementary approaches.
Our research on neurological noise is inspired by three main goals:

$\bullet$ \emph{To improve the performance of real-time neurological algorithms:} Certain neurological signal processing tasks, such as extracting event-related potentials (ERPs), increase  signal-to-noise ratios (SNRs) by averaging many epochs of data recorded over long experimental periods \cite{luck-book-2014}. However, time-frequency algorithms, which assume high levels of nonstationarity \cite{rankin-etal-2007}, and BCIs \cite{wangs-hong-gaos-2006,allison-et-al-2008}, which are meant to provide disabled patients with a sense of real-time control, must work with much shorter, single-trial epochs. 
	
	Ordinary linear filtering of such short epochs is problematic, since there is evidence that the brain's responses are highly nonlinear \cite{tetz-etal-2006}, and because the target signals can be nearly indistinguishable from the background (cf. \myfig{fig:RawEEG28Hz}). Moreover, poorly-fitting models of the detailed statistical characteristics of brain processes reduces the precision of detection/estimation procedures and makes model validation uncertain. 

	We are particularly interested in developing real-time SSVEP BCI algorithms which are accurate into the $\gamma$-band, a region of the EEG spectrum which is dominated by noise.

$\bullet$ \emph{To create statistically-realistic simulations of neurological signals:} The common practice is to add artificial noise sources to neurological simulations in order to increase their realism and to measure the performance of models and algorithms \cite{diba-lester-koch-2004}. Good simulations of neurological signals are essential for development and testing of medical and BCI algorithms. This is especially true for critical applications to human neurology \cite{tetz-etal-2006,rankin-etal-2007} for which experimentation is highly restricted. 

$\bullet$ \emph{To create new insight into the underlying neurological processes:}  Statistical models of neurological noise have had remarkable success in providing indirect tests of neuroscience hypotheses. This was spectacularly true for the elucidation of the acetycholine neurotransmission mechanisms  in the 1970's by Katz \& Miledi \cite{katz-miledi-1972}. But noise models continue to enhance our understanding of neurological illness \cite{diez-etal-2013, vysata-etal-2014}, cognitive processes \cite{legenstein-maass-2014}, and may even explain brain nonlinearity \cite{miller-troyer-2002}.
	 
	 We believe that the theory of neurological noise processes will play an increasingly important role in interpreting the behavior of the billions of neurons and trillions of ion channels in large-scale, biological brain networks.

\subsection{Steady-State Visual Evoked Potential BCIs}\label{subsec:SSVEP_def}

	Sudden stimuli such as a touch, a sound, or a bright flash will elicit a detectable brain reaction called an \myemph{event-related potential} \cite{luck-book-2014} or \myemph{evoked potential} (EP). Such EPs typically last on the order of 500 ms before disappearing, and usually may be reinvoked after a short refractory period. However, if the stimuli are repeated at a regular rate faster than about 2Hz, the EPs will not have time to decay and the brain's reaction will be a periodic signal called a \myemph{steady-state evoked potential} whose fundamental frequency is the same as the stimulus'. In particular, periodic visual stimuli will cause an SSVEP \cite{vialatte-maurice-dauwels-cichocki-2010}.
	
	These stimulus-dependent brain frequencies can be used to control BCIs \cite{allison-et-al-2008}, by flashing lights at various distinct frequencies simultaneously in different locations on a computer screen or LED device. The strongest SSVEP response peak detected corresponds to the location on which the subject's attention was most focused and usually represents the selection.
	
	Low frequency visual stimuli also induce harmonics \cite{herrmann-2001} (cf. \myfig{fig:chi2_8Hz}) which also may be used for BCI detection \cite{SSVEP-BCI-harmonics-2005}. Thus the corresponding subject selections can be identified by simple algorithms. It is worth remarking that such harmonic responses prove conclusively that the brain is a nonlinear system since pure linear systems cannot generate harmonics. On the other hand, higher frequencies, especially those which approach the gamma-band above 30 Hz, are much more difficult to detect, because their response power is close to that of the background and all harmonics (if they exist at all) and are lost in the gamma-band noise. This is seen most clearly in \myfig{fig:RawEEG28Hz}, which shows a 28 Hz brain response almost indistinguishable from background noise. (However, note that our new GVZM-based algorithms in \mysec{sec:SSVEP} detect the 28 Hz peak even in this difficult data set.)

\subsection{Modeling EEG Noise and Noise Power Spectral Densities}\label{subsec:SSVEP_SFT}

	EEG noise has often been modeled using  PSDs \cite{oppen-schafer-book-1975,brockwell-davis-book-1991} that are power law functions \cite{mandelbrot-book-1999a,miller-troyer-2002} of the form $S(f) \propto 1/f^\theta$ for some $0 \myle \theta < 2$. Neurological noise PSDs at all scales have long been claimed \cite{hill-chen-IV-1972,linkenhaer-hansen-etal-2001,diba-lester-koch-2004} to have the general characteristics of such ``$1/f$--type noise'' \cite{mandelbrot-book-1999a} even though this claim implies the obvious paradox of infinite total power in a biological system. 
	
	Also, autoregressive (AR) \cite{allison-et-al-2008} and autoregressive-moving average (ARMA) models of \cite{brockwell-davis-book-1991} of EEG recordings are very commonly used, in particular to simulate noise for the SSVEP detection algorithm \cite{liavas-moustakides-1998}. Such models are useful approximations but yield PSDs which are rational functions and are more appropriate to stationary, linear systems. However, brain signals are known to be neither stationary \cite{rankin-etal-2007} nor linear \cite{miller-troyer-2002}.

	A useful approach has been to model the statistical variations of noise periodograms around their mean PSD. In many cases, the periodogram values $S_x\Paren{k}$ of a discrete-time random process $x\Paren{n}$ are independent, scaled $\chi^2\Paren{2}$ random variables \cite{mood-graybill-boes-book-1974}, whose expected value $\Expv\Brack{S_x\Paren{k}}$ gives the mean spectral power at the frequency index $k$ \cite{brockwell-davis-book-1991}. This result holds exactly for many important special cases, such as white Gaussian noise and causal periodic ARMA processes. (See Appendix A for the definition.) In more general situations, the $\chi^2\Paren{2}$ distribution is approached only asymptotically as the data length increases, however the result still has very broad applicability \cite{brockwell-davis-book-1991}.

	The idea of detecting SSVEP responses for BCI applications by performing statistical testing of the periodogram (often called the Spectral \textit{F}-Test (SFT)) was developed by several research groups in the mid-90's \cite{sa-infantosi-lazarev-2007} and is based on ``hidden periodicity'' methods dating back to the 40's \cite{brockwell-davis-book-1991}. The idea has been used regularly, most notably in \cite{liavas-moustakides-1998,baka-tana-cich-neuro-let-2010} which enhanced the flexibility of the original SFT procedure.

\subsection{Contributions and Paper Organization}\label{subsec:Contributions}

The central innovation of this article is a new statistical model  of the discrete Fourier transform (DFT) periodograms \cite{oppen-schafer-book-1975} of EEG noise processes, the so-called GVZM noise model. We validate this model via three distinct approaches: (i) developing a neurological noise model based on quantum mechanical ion channel kinetics, (ii) linking autoregressive time series to the GVZM noise model and EEG noise, (iii) and designing and evaluating two real-time SSVEP BCI estimation algorithms, based on the new GVZM noise model. Our contributions can be summarized as the following

$\bullet$ We introduce a novel, five-parameter GVZM PSD function, and we show that this function accurately matches the PSD function of recorded EEG noise, from near 0 Hz to 30 Hz.  Also, we show that this function is approximately of the form $1/f^\theta$ in the mid-frequencies but has finite total power and finite amplitude as $f \to 0$.  We provide a biophysical meaning of the function parameters and indicate how the function can be derived from a theory of ion channel noise. 

$\bullet$ We use the GVZM PSD function to define
a new statistical model of the DFT periodgrams of EEG noise processes, which we call the GVZM noise model. We show that this model successfully captures the statistics of EEG periodograms, allows rigorous statistical testing, detection, and estimation, and enables statistically accurate simulations of EEG noise periodograms. 

$\bullet$ We define a class of mixed autoregressive (AR) time series, which we call AR-GVZM processes, containing accurate simulations of EEG noise time series, and whose PSD functions converge asymptotically to the GVZM PSD function (GVZM noise model). We show also that, if recorded EEG noise is assumed to be Gaussian and satisfies the GVZM noise model, it can be approximated arbitrarily closely by AR-GVZM processes.

$\bullet$ We define two new SSVEP frequency estimation algorithms based on the GVZM noise model: a simple $\chi^2$-distribution detection algorithm for frequency spikes and a more sophisticated, $F$-distribution estimator based on the well-known algorithm in \cite{liavas-moustakides-1998}. We demonstrate statistically that these algorithms outperform two well-known rivals \cite{vialatte-maurice-dauwels-cichocki-2010,wei-xiao-lu-2011,kus-dusyk-etal-2013}, thus providing evidence for the truth of our noise model. The practical success of the GVZM noise model provides indirect validation of the underlying ion channel noise model from which it was developed.

The remainder of this paper is organized as follows: \mysec{sec:GVZM}  introduces the GVZM PSD function, defines the GVZM noise model, and discusses the behavior of the GVZM PSD function. \mysec{sec:Channels} briefly summarizes the results of the theory of quantum-controlled ion channels. \mysec{sec:ARGVZM} introduces AR-GVZM processes and discusses their link to GVZM noise model and EEG noise. \mysec{sec:SSVEP} describes and compares statistically two GVZM-based SSVEP algorithms and two rival SSVEP estimators. \mysec{sec:Summary} presents our conclusions.


\begin{figure} [h!]
\centering
\vspace{-.02\textwidth}
\includegraphics[width=0.45\textwidth]{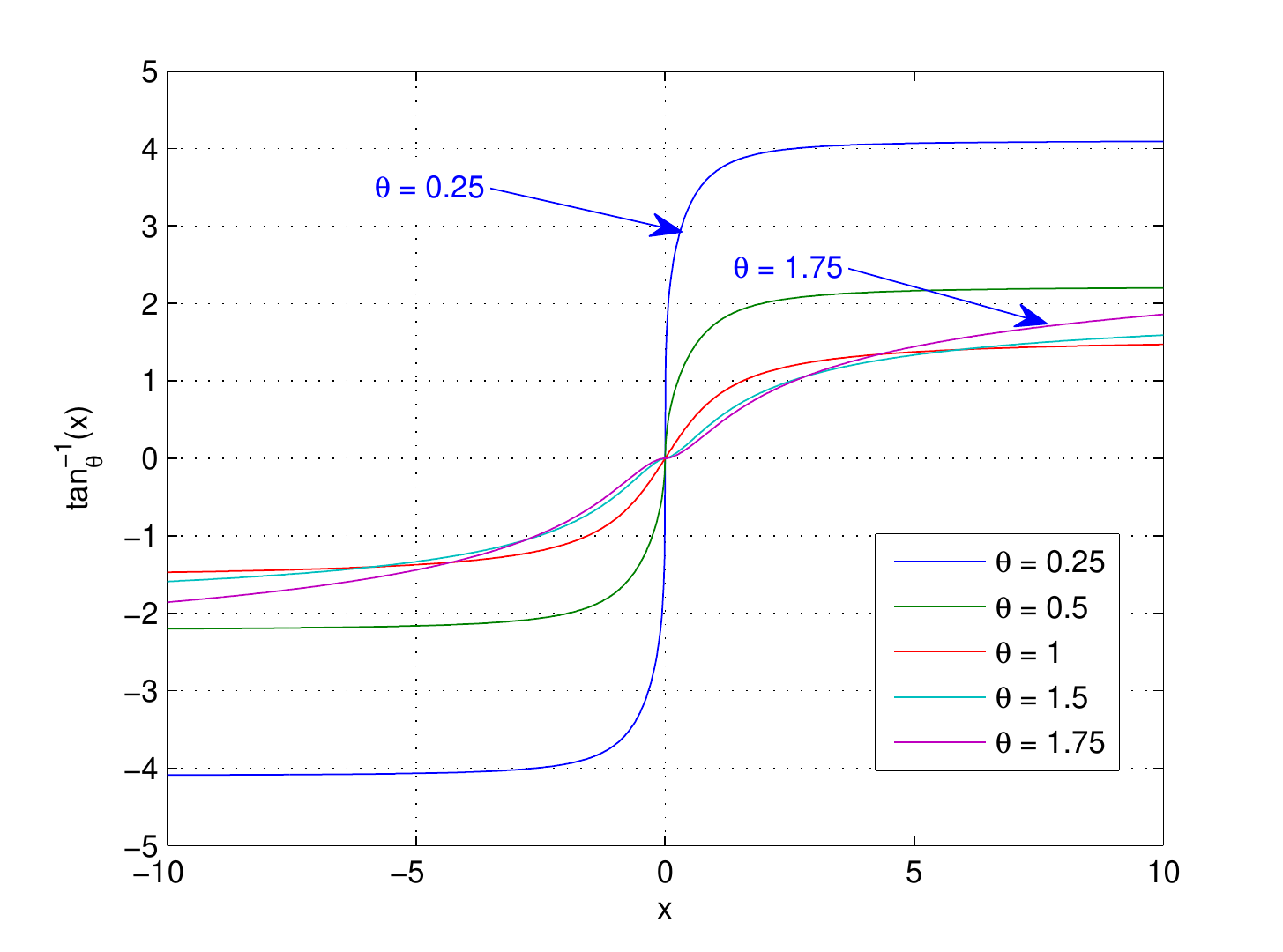}
\vspace{-.02\textwidth}
\caption{The generalized arctangent function $\mathrm{tan}_\theta^{-1}\Paren{x}$.}
\label{fig:atan}
\vspace{-.03\textwidth}
\end{figure}

\section{The GVZM EEG Periodogram Model}\label{sec:GVZM}

	In this section, we first specify the notion of periodogram we are using and then define the GVZM PSD function. We then present the GVZM EEG periodogram model.

\subsection{Defining the Periodogram}\label{subsec:Periodograms}

\mybegdef{def:periodogram}
	For the purpose of this article, the single-epoch \myemph{periodogram} of a discrete-time signal $x\Paren{n}$, $0 \myle n \myle N-1$, is defined as
\[
 	S_x\Paren{k} \Eqdef \frac{2\pi}{N} \Abs{X\Paren{k}}^2, \;\; 0 \myle k \myle N-1,
\]

\noindent where $X\Paren{k} \Eqdef \sum_{n = 0}^{N-1} {x\Paren{n} \; \Exp{-i \Paren{2\pi/N}kn } }  $
is the DFT of $x\Paren{n}$ \cite{oppen-schafer-book-1975}. The factor $2\pi$ is present to convert $S_x\Paren{k}$ to dimension amplitude$^2$/frequency, where frequency has dimension 1/time, not angle/time; i.e., units of Hz rather than radians/sec.
\myenddef


\begin{figure*}
	\vspace{-.02\textwidth}
	\centering
	\begin{subfigure}[t]{0.45\textwidth}
		\includegraphics[width=\textwidth]{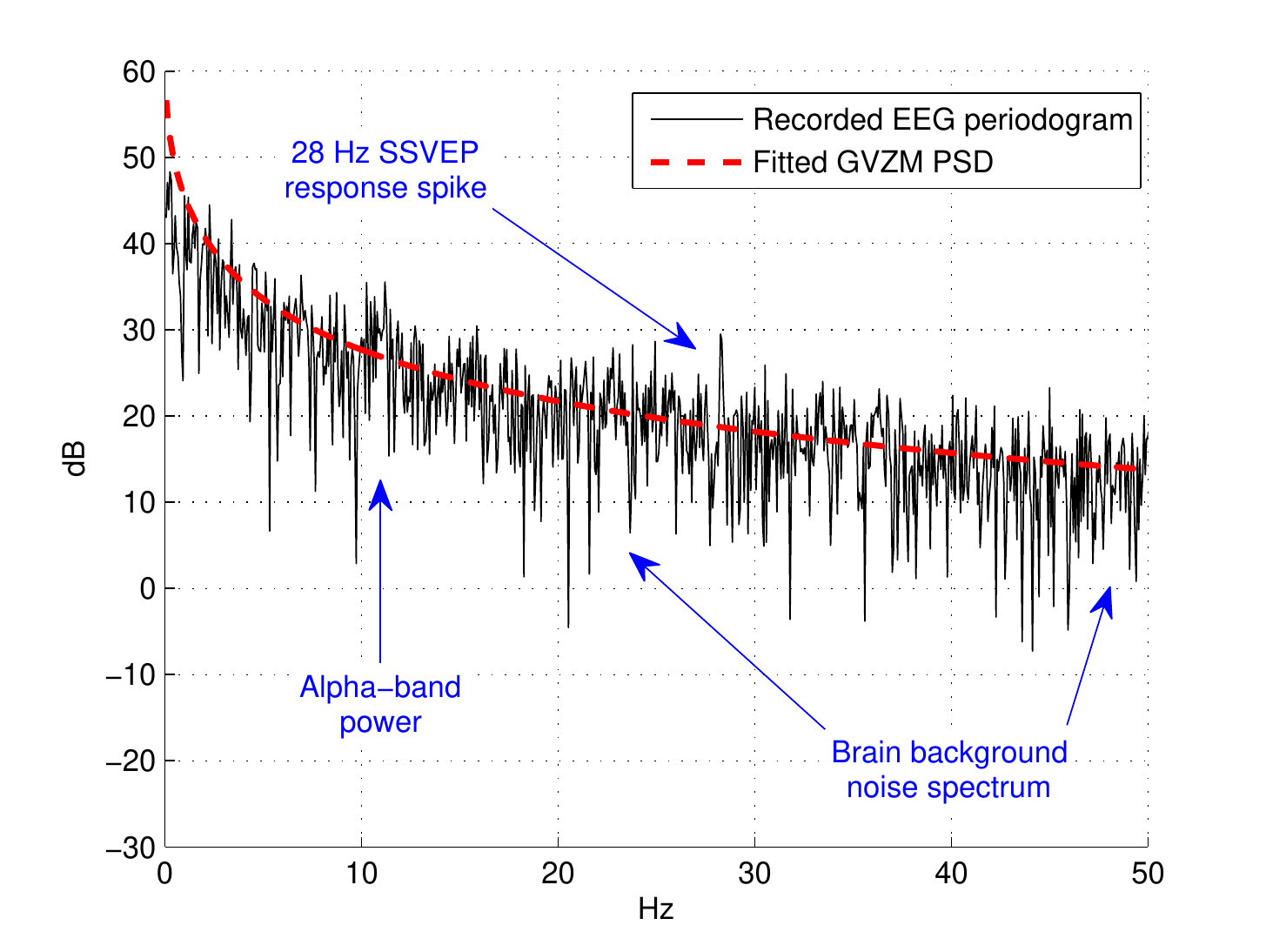}
		\caption{}
		\label{fig:GVZMfit}
	\end{subfigure}
	\hfill
	\begin{subfigure}[t]{0.45\textwidth}
		\includegraphics[width=\textwidth]{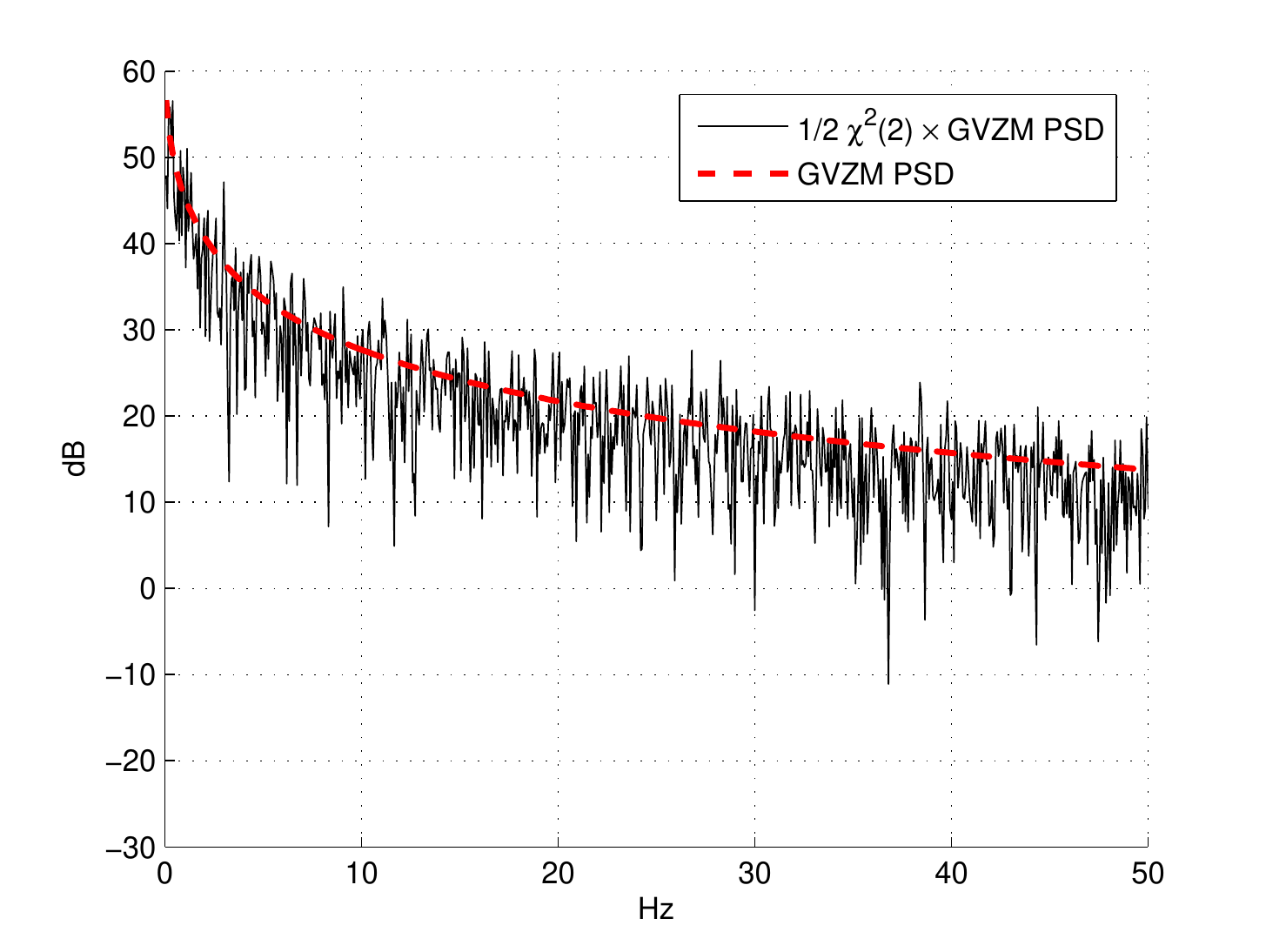}
		\caption{}
		\label{fig:chi2_simulation}
	\end{subfigure}
	\hfill
	\vspace{-.02\textwidth}
	\caption{(a) $\mathrm{GVZM} $ PSD function fitted to the 28 Hz stimulus recorded EEG periodogram of \myfig{fig:RawEEG28Hz}. (b) $\mathrm{GVZM} \cdot \Paren{1/2} \chi^2\Paren{2} $ simulated EEG periodogram using the parameters derived from (a).}\label{}
	\vspace{-.04\textwidth}
\end{figure*}



\subsection{Defining the GVZM PSD Function}\label{subsec:GVZM_def}

\mybegdef{def:GVZM_atan}
	The \myemph{GVZM PSD}  $S_\mathrm{GVZM}\Paren{f}$ is the family of functions with five parameters $0 < \theta < 2$, $0 < \upsilon_1 < \upsilon_2$, $P_0 \myge 0$, $P_\mathrm{s} \myge 0$ and defined by
\begin{align}\label{eq:SVZM_def}
	S_\mathrm{GVZM}\Paren{f} \Eqdef P_0 \Abs{f}^{-\theta}
		&\Paren{\mathrm{tan}^{-1}_\theta\Paren{2\pi \upsilon_2 \Abs{f}} \right. \nonumber\\
			&\quadOne\left. - \mathrm{tan}^{-1}_\theta\Paren{2\pi \upsilon_1 \Abs{f}} } + P_\mathrm{s}, 
\end{align}
where
\mybegeq{eq:atan_def}
	\mathrm{tan}^{-1}_\theta\Paren{x} \Eqdef \mathrm{sgn}\Paren{x} \int\limits_0^{\Abs{x}} 
		{   \frac{u^{\theta-1}}{1+u^2}   du}.
\myendeq
\myenddef

Note that for $\theta = 1$, Eq. \myeq{eq:atan_def}  is the ordinary $\mathrm{arctan}\Paren{x}$ (cf. \myfig{fig:atan}). The dimension of $\upsilon_1$, $\upsilon_2$ is time while that of $P_0$, $P_\mathrm{s}$ is amplitude$^2$/frequency (i.e., noise power). The spectral exponent $\theta$ is dimensionless. The biophysical meaning of these parameters is described in \mysec{subsec:QuantumChannels}.
	
	The importance of this definition is that, so far as the authors are aware, this is the first simple model of the average EEG background noise spectrum proposed in the literature which can match recorded EEG periodograms from near 0 Hz to over 30 Hz, with a fixed number of parameters. Moreover, the GVZM PSD function approximates a power law  $1/f^\theta$ in the mid-frequencies without requiring infinite power. In fact, $S_\mathrm{GVZM}\Paren{f}$ always has finite amplitude and finite total power.
	
	In the next section, the GVZM PSD will be used to define a statistical model of EEG periodograms, the core of our work and the basis of the new SSVEP algorithms in \mysec{sec:SSVEP}.

\subsection{The GVZM Noise Model}\label{subsec:GVZM_model}

\mybegdef{def:GVZM_model} Let $x\Paren{n}$, $0 \myle n \myle N-1$ be samples of an EEG electrode, taken at sampling frequency $f_\mathrm{samp}$, which measures a naturally-occuring, stationary ensemble of brain background processes for a single individual. Then the \textbf{GVZM noise model} of the periodogram values $S_x\Paren{k}$, $0 \myle k \myle N-1$ is the random process
\mybegeq{eq:GVZM_process}
	S_x\Paren{k} = S_\mathrm{GVZM}\Paren{ k \Delta f} \cdot \Paren{1/2} \, \Xi\Paren{k},
\myendeq
\noindent where $\Xi\Paren{k}$ is  a sequence of $\chi^2\Paren{2}$ distributed \cite{mood-graybill-boes-book-1974} random variables, which are independent for $0 \myle k, l < N/2$ when $k \ne l$,  $\Delta f \Eqdef f_\mathrm{samp} / N$, and $S_\mathrm{GVZM}\Paren{f}$ is an appropriate GVZM PSD defined by \myeq{eq:SVZM_def}.
\myenddef

	Note that the reason for the restriction of independence to $0 \myle k, l < N/2$ is that, since $x\Paren{\Hole}$ is real, we have $S_x\Paren{N-k} = S_x\Paren{k}$. Also note that the half-interval definition $0 \myle k, l < N/2$ applies whether $N$ is even or odd.

	We can write \myeq{eq:GVZM_process} informally as
\[
	S_x\Paren{k} \sim  S_\mathrm{GVZM}\Paren{ k \Delta f} \cdot \Paren{1/2} \, \chi^2\Paren{2},
\]
where $\sim$ denotes ``is distributed as.''

	The $\Paren{1/2} \, \chi^2\Paren{2}$ periodogram distribution holds exactly for special processes such as $N$-periodic ARMA described in Appendix A. When conditions for the Central Limit Theorem hold, the $\Paren{1/2} \, \chi^2\Paren{2}$ distribution holds asymptotically as the data length $N \to \infty$ \cite{brockwell-davis-book-1991}. Both $N$-periodic ARMA and the large-$N$ approximations are appropriate for EEG noise.

	Note that Eq. \myeq{eq:GVZM_process} implies that the expected values $\Expv \Brack{S_x\Paren{k}}$ are equal to $S_\mathrm{GVZM}\Paren{ k \Delta f}$. Using general results presented in \cite{brockwell-davis-book-1991}, the converse holds asymptotically; that is, the process $S_x\Paren{k}$ converges uniformly in distribution to $S_\mathrm{GVZM}\Paren{ k \Delta f} \cdot \Paren{1/2} \, \Xi\Paren{k}$ as  $N \rightarrow \infty$, at least when conditions for Central Limit Theorem hold. (cf. \cite{brockwell-davis-book-1991}, Chapter 10 for details.)

	\mydef{def:GVZM_model} is consistent with the single-epoch approach of \cite{brockwell-davis-book-1991,liavas-moustakides-1998} but multi-epoch averages of such single-epoch spectra can be used as well. Our methods will apply to such general periodograms merely by replacing ``$\chi^2\Paren{2}$'' with ``$1/M \; \chi^2\Paren{2M}$,'' where $M$ is the number of (statistically independent) epochs.

\subsection{The Behavior of the GVZM PSD  Function}\label{subsec:GVZM_behavior}

	The properties of $\mathrm{tan}^{-1}_\theta\Paren{x}$ show that once the noise floor $P_\mathrm{s}$ is subtracted  we have the approximate proportionalities $S_\mathrm{GVZM}\Paren{f} \propto1/f^0$ (i.e., a constant) for $f < 1/\Paren{2\pi \upsilon_2}$, $S_\mathrm{GVZM}\Paren{f} \propto 1/f^\theta$ for $1/\Paren{2\pi \upsilon_2} < f < 1/\Paren{2\pi \upsilon_1}$, and  $S_\mathrm{GVZM}\Paren{f} \propto 1/f^2$ for $f > 1/\Paren{2\pi \upsilon_1}$. Thus its roll-off transitions smoothly through the $1/f^\theta$ regime, without any of the so-called ``catastrophes'' \cite{mandelbrot-book-1999a} of apparent infinite power when $f \to 0$ and infinite integrals as $f \rightarrow \infty$ which plague true $1/f$--type noises. In particular,  $S_\mathrm{GVZM}\Paren{f}$ approaches a (calculable) limiting value as $f \rightarrow 0$.

	\myfig{fig:GVZMfit} shows a GVZM PSD function fitted to the periodogram of recorded EEG data from an SSVEP session. The GVZM curve follows the periodogram closely except for the alpha-band power \cite{nunez-srin-book-2006} and the SSVEP response spike. 
	
	 \myfig{fig:GVZMfit} should be compared closely to \myfig{fig:chi2_simulation} which shows a \myemph{simulated} periodogram based on the fitted  GVZM PSD function and \myeq{eq:GVZM_process}; that is, samples of an independent and identically distributed (iid) $\chi^2\Paren{2}$ pseudo--random process $\Xi\Paren{k}$, $0 \myle k \myle N-1$ were generated and each sample was multiplied by the factor $\Paren{1/2}\cdot \, S_\mathrm{GVZM}\Paren{ k \Delta f}$. The results were then plotted against frequency on a log-linear scale. We observe that the simulated periodogram of \myfig{fig:chi2_simulation} is visually indistinguishable from the background periodogram of \myfig{fig:GVZMfit}. 
	
	Eq. \myeq{eq:SVZM_def} has it origin in investigations dating to the 1930's \cite{bernamont-1937} on apparent $1/f$-type noise in vacuum tubes and semiconductors. In 1957, A.L. McWhorter proposed\cite{mcwh-1957} a simple explanation for $1/f$--type semiconductor noise at thermal equilibrium, based on the assumption that the logarithm of the rate at which electrons drop from an activated state was proportional to the energy of that state. Subsequently A. van der Ziel and others \cite{vdziel-1978} abstracted the McWhorter mechanism to general noise processes. Our formula \myeq{eq:SVZM_def} reduces to the original van der Ziel-McWhorter PSD function for $\theta = 1$.

\section{Neurological Noise Theory and Quantum Mechanical Ion Channel Kinetics}\label{sec:Channels}

	In this section we briefly discuss the theoretical foundation of our GVZM PSD function in \mydef{def:GVZM_atan}, using a new model of \myemph{ion channel quantum stochastic processes} we have developed, and whose details are presented in the supplementary document \cite{paris-atia-vosoughi-berman-2015a}. For the present paper we extract a simplified form of one type of potential autocorrelation function in \cite{paris-atia-vosoughi-berman-2015a} and provide the biophysical interpretation of its parameters.

\subsection{Are Ion Channels the Source of \textit{1}$/f$--type Noise?}\label{subsec:NoisyChannels}

	Ion channels are protein-based micromachines densely embedded in all neuron membranes, which create and control the transmission of information by regulating the passage of ions in response to neurotransmitters, local voltages, or external stimuli such as temperature, pressure, or photon reception \cite{hille-book-2001}. 
	
	Soon after Hodgkin \& Huxley decoded the generation of action potentials by the $\Kion$/$\Naion$ channel system \cite{hodgkin-huxley-1952}, researchers began to model ion conductance variations in membranes as resulting from simple Markov processes governing the open/close kinetics of the embedded ion channels \cite{hill-chen-IV-1972}. These conductance variations were recognized quickly as a potential source of neurological noise, at least at the neuron level \cite{hille-book-2001}.
	
	In a series of articles \cite{hill-chen-IV-1972} in the early 1970's, Hill \& Chen examined several versions of Markov models, including one which was claimed to test the original van der Ziel-McWhorter paradigm (\mysec{sec:GVZM}), with the stated goal of \myemph{excluding} ion channels as sources of the $1/f$--type noise component seen in neuron-level recordings. The limited computational resources available at that time prevented detailed simulations, and they eventually rejected ion channel kinetics as the source of the $1/f$--type noise, thus leaving its origin unexplained. Since then there have been attempts to explain this neurological noise by various other mechanisms \cite{diba-lester-koch-2004}, however, none of these explanations have been completely successful.

\subsection{Maximum Entropy in Populations of Quantum Ion Channels and GVZM Noise}\label{subsec:QuantumChannels}
	
	We have revisited the conclusion of of Hill \& Chen in \cite{paris-atia-vosoughi-berman-2015a}  by considering network-level behavior of large populations of ion channels in statistical equilibrium. Specifically, we have returned to McWhorter's original model (\mysec{sec:GVZM}) of noise generation in semiconductors and identified channel analogies to McWhorter's energy-dependent rate constants and maximum entropy distributions. We have applied McWhorter's abstract maximum entropy paradigm by creating a new quantum mechanical model of ion channel kinetics, which we call \myemph{activated measurement processes}. In the following, we recall a special case of our work in \cite{paris-atia-vosoughi-berman-2015a}:
	
\mybegthm{thm:ionR}

	Let $C$ be a population of ion channels whose kinetic rate matrices at gating level $V$ are obtained from activation energy operators $\Curly{\Op{E}_c\Paren{V} \st c \in C}$ according to the standard activated measurement model \cite{paris-atia-vosoughi-berman-2015a}. Let $\Angle{\Op{\Psi}\Paren{c},p\Paren{c} \st c \in C}$ be the entropy-maximizing state and density functions under the energy constraint
\[
	\int\limits_C {\Trace\Brack{\Op{E}_c\Paren{V} \cdot \Op{\Psi}\Paren{c}} \, p\Paren{c} dc} = \bar{E}
\]
for fixed gating level $V$ and mean activation energy $\bar{E}$, where $\Trace\Brack{\Hole}$ is the trace operator.

	Then there are conditions such that the autocorrelation of the population's channel conductance $g$ is 
\mybegeq{eq:R_GVZM}
	R_g\Paren{\tau} = P_0 \int\limits_{\upsilon_1}^{\upsilon_2}{\frac{1}{\upsilon^{2-\theta}}  \Exp{-\Abs{\tau}/\upsilon} d\upsilon} + P_\mathrm{s} \cdot \delta\Paren{\tau}, 
\myendeq
for constants $P_0$,$\upsilon_1$,$\upsilon_2$,$\theta$,$P_\mathrm{s}$.

\myendthm

(The ``conditions'' mentioned in \mythm{thm:ionR} are technical and are omitted.)

Note that the Fourier transform of \myeq{eq:R_GVZM} is easily seen to be the GVZM PSD function in \mydef{def:GVZM_atan}.

	While these results have immediate relevance only for populations of ion channels, we have argued in \cite{paris-atia-vosoughi-berman-2015a} that the presence of GVZM noise in large-scale EEG recordings ultimately derives from the collective behavior of billions of ion channels balanced in statistical equilibrium.

	A rough idea of the biophysical meaning of the parameters can be given. The values $\upsilon\Paren{x}$ represents the range of possible reciprocal eigenvalues (``modes'') of the kinetic rate matrices \cite{hille-book-2001} of all possible ion channels in a large network while $X$ is an index set for these modes. The $P_\mathrm{s} \cdot \delta\Paren{\tau}$ term represents ``sputter'' in the channels: the possibility that a channel's state is not quite open or closed completely. The $1/\upsilon\Paren{x}$ prefactor in the integrand derives from the proportionality between activation energy and $\mathrm{log}\Paren{\upsilon}$, a feature common to both the McWhorter paradigm as well as an alternative ion channel mechanism called Eyring rate theory \cite{woodbury-1969}.
	
	Most importantly, the exponent $\theta$ is the normalization parameter for a standard thermodynamic partition function 
\[
	Z\Paren{\theta} = \int\limits_X {\Exp{-\theta E\Paren{\upsilon}} \,  d\upsilon},
\]
where, as previously mentioned, the activation energy of mode $\upsilon$ is $E\Paren{\upsilon} \propto \textrm{log}\Paren{\upsilon}$. 

	This latter point is very significant. Many attempts have been made to explain irrational exponents in $1/f^\theta$--type spectra by means of purported power laws, ``scaling'', or ``self-similarity'' phenomena in neural tissue \cite{mandelbrot-book-1999a}. We consider our interpretation of spectral exponents as normalizing constants for logarithmically-determined energy levels to represent a needed influx of reality into the field of neurological noise.


\begin{figure} [h!]
\centering
\vspace{-.02\textwidth}
\includegraphics[width=0.45\textwidth]{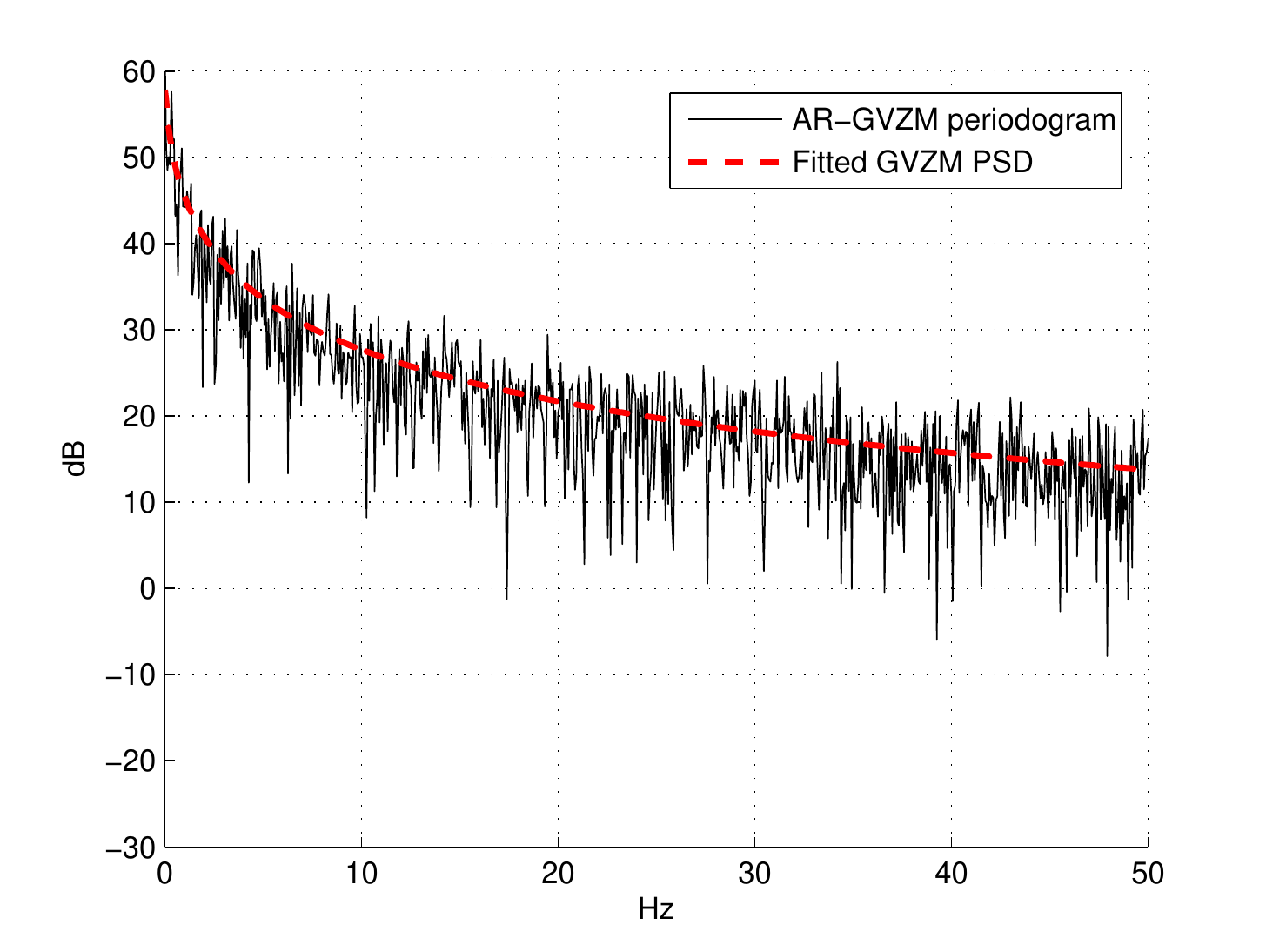}
\vspace{-.02\textwidth}
\caption{Periodogram of AR-GVZM simulation of EEG noise, with $K = 300$, using the \myfig{fig:GVZMfit} parameters.}
\label{fig:ARGVZM}
\vspace{-.04\textwidth}
\end{figure}

\section{Autoregressive GVZM Simulations of EEG Noise}\label{sec:ARGVZM}

	The GVZM noise model in \mydef{def:GVZM_model} concerns periodograms and makes no claims about the underlying time domain signal from which they were calculated. In this section, we define a class of mixed autoregressive time series called \textbf{autoregressive GVZM (AR-GVZM) processes}, which are useful simulations of EEG noise and are closely associated with GVZM periodograms. In \mysec{subsec:ARGVZMdef}, we define AR-GVZM processes and show an example based on the parameters derived from recorded data (\myfig{fig:ARGVZM}), in \mysec{subsec:ARConvergence} we connect the asymptotic properties of AR-GVZM, periodograms to the GVZM noise model.

\subsection {AR-GVZM, Processes} \label{subsec:ARGVZMdef}

	Let $0 < \upsilon_1 < \upsilon_2$ be fixed parameters with the dimension of time, $0 < \theta < 2$ be dimensionless, and $P_0,P_\mathrm{s} \myge 0$ have dimension amplitude$^2$/frequency. Let $K > 0$ be an integer and define $\Delta \upsilon = \Paren{\upsilon_2-\upsilon_1}/\Paren{K-1}$. Let $\Delta t > 0$ be a time step.
	
	For $0 \myle k \myle K-1$, define the AR coefficients $a_k,\,b_k$ and weights $w_k$ by
\mybegeq{eq:ARcoeffs}
\left\{
\begin{aligned}
a_k &\Eqdef \Exp{-\Delta t / \Paren{\upsilon_1 +k\Delta\upsilon}}, 
\quadTwo b_k\Eqdef \sqrt{1 - a_k^2}, \\
w_k &\Eqdef  \sqrt{\frac{1}{\Paren{\upsilon_1 +k\Delta\upsilon}^{2-\theta}} \, \Delta\upsilon \Delta t}.
\end{aligned}
\right.
\myendeq
	
	Let $x_k\Paren{n}$ be the \First\!--order, stationary AR process \cite{brockwell-davis-book-1991} 
\begin{align*}
	x_k \Paren{n} &= a_k\, x_k \Paren{n-1} + b_k\,  \sqrt{P_0} \, e_k\Paren{n},
\end{align*}
where the error processes $e_k\Paren{n}$ are iid $N\Paren{0,1}$ random variables.

	Then EEG background processes can be modeled by the discrete-time, mixed AR simulations
\begin{align}\label{eq:ARGVZM}
	x_K\Paren{n} &\Eqdef \sum\limits_{k=0}^{K-1} {w_k} {\cdot x_k\Paren{n}} 
			+ \sqrt{P_\mathrm{s}} \cdot f_\mathrm{s}\Paren{n},
\end{align}
where the error process $f_\mathrm{s}\Paren{n}$ is an $N\Paren{0,1}$ random variable which is independent of all the $e_k\Paren{n}$'s.
	We call such processes AR-GVZM, simulations of EEG noise.

	\myfig{fig:ARGVZM} shows the periodogram of such a simulation $x_K\Paren{n}$, with $K = 300$ AR subprocesses $x_k\Paren{n}$, using the GVZM parameters that optimally fit the data of \myfig{fig:GVZMfit}. It can be seen how accurately the periodogram of $x_K\Paren{n}$ matches the characteristics of the recorded data, except for the SSVEP response spike and the excess alpha-band power. Note that \myfig{fig:ARGVZM} is the periodogram of a simulated EEG time series, while \myfig{fig:chi2_simulation} shows the direct simulation of an EEG periodogram, with no underlying time series.

\subsection{The Convergence of AR-GVZM, Processes to the GVZM Noise Model}\label{subsec:ARConvergence}

	We choose some $0 \myle k \myle K-1$ and let $\upsilon =\upsilon_1 +  k\Delta\upsilon$ in \myeq{eq:ARcoeffs}. For small $\Delta t$, we find $a_k = 1-\Delta t / \upsilon$ and $b_k = \sqrt{2\Delta t / \upsilon}$, where we have ignored terms of order higher than $\Delta t$. Defining $\Delta x_k\Paren{n} \Eqdef x_k\Paren{n}-x_k\Paren{n-1}$, we easily calculate
\mybegeq{eq:dx-dt}
	\frac{\Delta x_k\Paren{n}}{\Delta t} + \frac{1}{\upsilon}\cdot x_k\Paren{n-1} =
		\sqrt{\frac{2 P_0}{\upsilon \Delta t}}\cdot e_k\Paren{n}.
\myendeq
	
	Let $y_k\Paren{n} \Eqdef w_k\cdot x_k\Paren{n}$. Using the definition of $w_k$ in \myeq{eq:ARcoeffs}, \myeq{eq:dx-dt} becomes
\[
	\frac{\Delta y_k\Paren{n}}{\Delta t} + \frac{1}{\upsilon}\cdot y_k\Paren{n-1} = 
		\sqrt{\frac{2P_0}{\upsilon\cdot\upsilon^{2-\theta}}\Delta\upsilon}\cdot e_k\Paren{n}.
\]
	Now suppose the values $e_k\Paren{n}$ are samples $e_k\Paren{n\Delta t}$ of a continuous-time, iid $N\Paren{0,1}$ process $e_k\Paren{t}$. Then, as $\Delta t \to 0$, $y_k\Paren{n}$ will approach a continuous-time, stationary, Gaussian AR process $y_k\Paren{t}$ satisfying the stochastic differential equation
\mybegeq{eq:AR_ODE}
	\frac{dy_k}{dt} + \frac{1}{\upsilon}\cdot y_k\Paren{t} = 
		\sqrt{\frac{2P_0}{\upsilon\cdot\upsilon^{2-\theta}}\Delta\upsilon}\cdot e_k\Paren{t}.
\myendeq
	It is well-known \cite{dest-book-2012} that the autocorrelation function of $y_k\Paren{t}$ satisfying \myeq{eq:AR_ODE} is
\[
	R_k\Paren{\tau} = P_0\,\frac{1}{\upsilon^{2-\theta}}\Exp{-\Abs{\tau}/\upsilon} \Delta\upsilon.
\]
	Using \myeq{eq:ARGVZM}, suppose the values $f_\mathrm{s}\Paren{n}$ are also samples $f_\mathrm{s}\Paren{n\Delta t}$ of a continuous-time, iid $N\Paren{0,1}$ process $f_\mathrm{s}\Paren{t}$ which is independent of all the $e_k\Paren{t}$'s. We observe that the AR-GVZM, processes $x_K\Paren{n}$ in \myeq{eq:ARGVZM} approach a continuous-time, stationary, Gaussian, mixed AR process with autocorrelation function
\[
\begin{aligned}
	R_K\Paren{\tau} = 
	P_0\,\sum\limits_{k=0}^{K-1}  \frac{1}{\upsilon^{2-\theta}}\Exp{-\Abs{\tau}/\upsilon}\Delta\upsilon
			+ P_\mathrm{s} \cdot \delta\Paren{t}.
\end{aligned}
\]
Therefore, as $K \to \infty$, the $x_K\Paren{t}$'s themselves will approach a Gaussian process $x_\ast\Paren{t}$ with autocorrelation 
\mybegeq{eq:R_ARGVZM}
	R_\ast\Paren{\tau} = P_0 \int\limits_{\upsilon_1}^{\upsilon_2}{\frac{1}{\upsilon^{2-\theta}}  \Exp{-\Abs{\tau}/\upsilon} d\upsilon} + P_\mathrm{s} \cdot \delta\Paren{\tau}.
\myendeq
It is now easy to check that the PSD function of $x_\ast\Paren{t}$, which is the Fourier transform of $R_\ast\Paren{\tau}$ \cite{oppen-schafer-book-1975},  is given precisely by the formula of Eq. \myeq{eq:SVZM_def}; i.e., the GVZM PSD function.

	This is a significant result since zero-mean, Gaussian processes are uniquely defined by their autocorrelation function \cite{brockwell-davis-book-1991}. Noting that the GVZM periodogram model implies the autocorrelation must be given by \myeq{eq:R_ARGVZM}, if we make the additional assumption that the EEG noise time series is zero-mean, Gaussian, then that time series \myemph{must} be $x_\ast\Paren{t}$. This means that the AR-GVZM, simulations can be made to approximate EEG background noise with arbitrary precision by means of the double limiting process $\Delta t, \Delta\upsilon \to 0$ described above.
	
	Note also that the converse can be proven using methods of \cite{brockwell-davis-book-1991}; that is, if the EEG noise time series is given by limits of AR-GVZM, processes, then the GVZM periodogram model must be valid.  This is important for practical applications because it defines the correct statistical model of the periodogram if AR-GVZM,  approximations are used to simulate EEG noise.

\section{Real-time BCI Algorithms and EEG Noise}\label{sec:SSVEP}

	In this section, we explore two new SSVEP frequency estimation algorithms we have designed, based on the GVZM noise model, which we refer to as \textbf{GVZM-}$\boldsymbol{\chi^2}$ and the \textbf{GVZM-}$\boldsymbol{F}$. We evaluate the performance of the new algorithms by comparing each with an existing, commonly-used procedure (to be be described later). Our statistical analysis demonstates that the GVZM-based algorithms outperform both their rivals.

\begin{figure*}
	\vspace{-.02\textwidth}
	\centering
	\begin{subfigure}[t]{0.47\textwidth}
		\includegraphics[width=\textwidth]{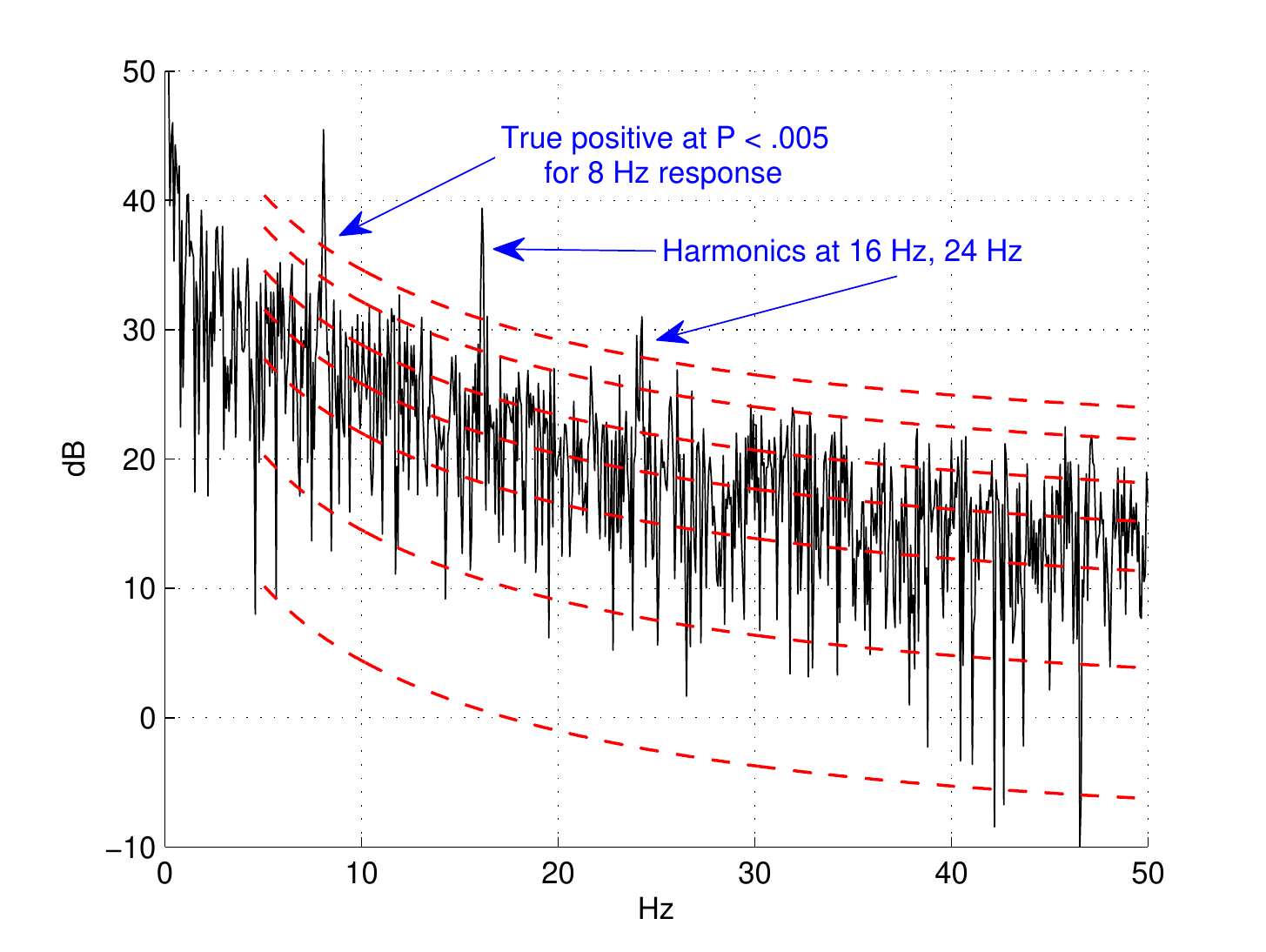}
		\caption{}
		\label{fig:chi2_8Hz}
	\end{subfigure}
	\hfill
	\begin{subfigure}[t]{0.47\textwidth}
		\includegraphics[width=\textwidth]{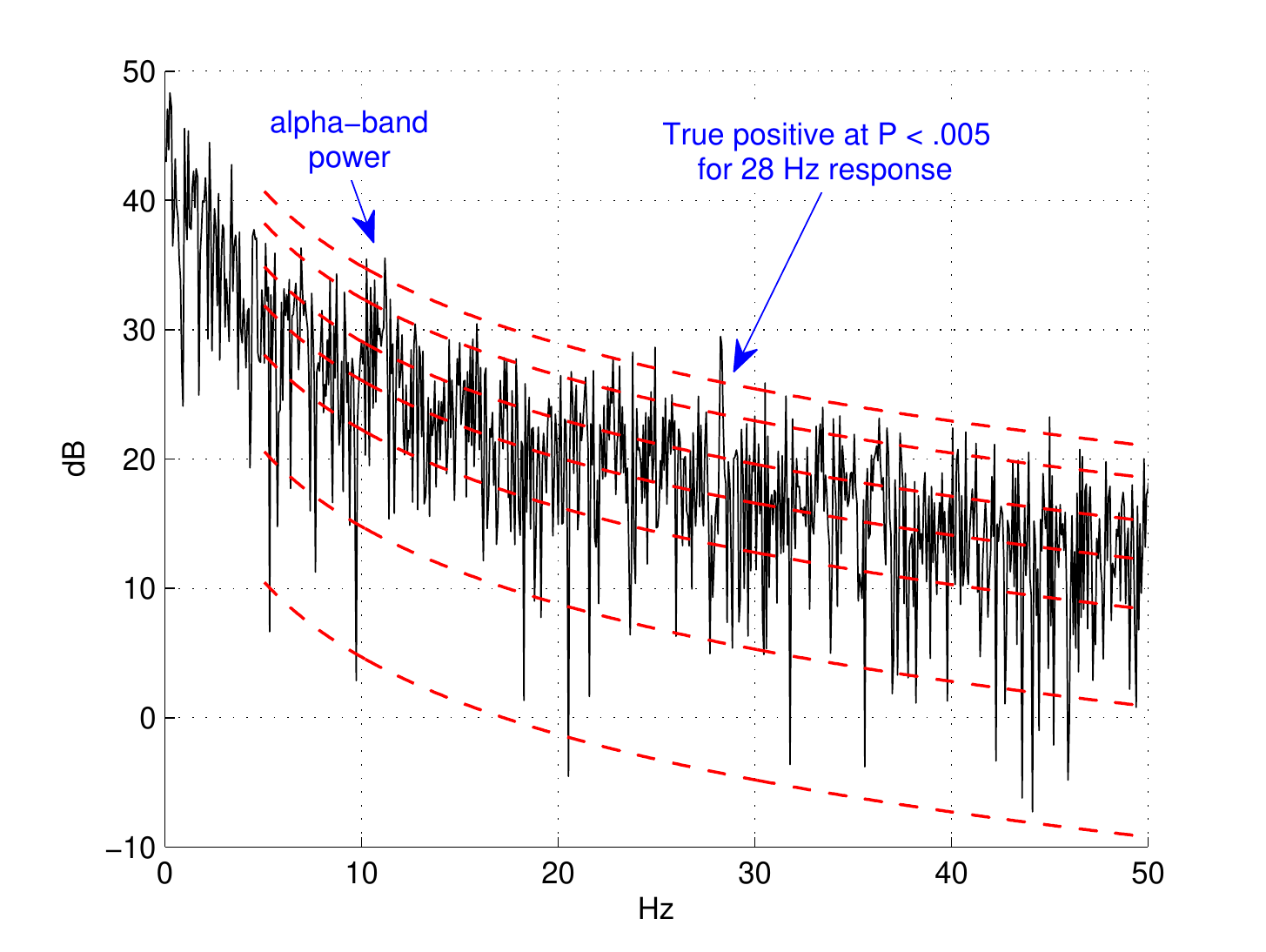}
		\caption{}
		\label{fig:chi2_28Hz}
	\end{subfigure}
	\hfill
	\vspace{-.01\textwidth}
	\caption{$\mathrm{GVZM} \cdot \Paren{1/2} \, \chi^2\Paren{2}$ critical levels corresponding to \textit{P} = 0.005, 0.05, 0.25, 0.5, 0.75, 0.95, 0.995 (top to bottom). (a) 8 Hz stimulus. (b) 28 Hz stimulus.}\label{}
	\vspace{-.04\textwidth}
\end{figure*}

	In \mysec{subsec:GVZM-Chi2} we examine the GVZM-$\chi^2$ algorithm, which is based on GVZM-$\chi^2$\textbf{-critical levels} for the EEG periodogram. These critical levels are curves drawn on the graph of the periodogram, which are parallel to the GVZM spectrum $S_\mathrm{GVZM}\Paren{f}$ defined in \mysec{subsec:GVZM_def}. Each represents the PSD level beneath which a random $S_\mathrm{GVZM}\Paren{f}\cdot\Paren{1/2}\chi^2\Paren{2}$ variable should remain, with specified probability (see \myfig{fig:chi2_8Hz} and \myfig{fig:chi2_28Hz} for examples of GVZM-$\chi^2$-critical levels.)
	
	In \mysec{subsec:SNR}, we perform a statistical performance comparison of  GVZM-$\chi^2$ against a commonly-used BCI algorithm \cite{wangs-hong-gaos-2006,wei-xiao-lu-2011,kus-dusyk-etal-2013} we refer to as \textbf{BCI-SNR}. Note that BCI-SNR is based on forming certain ratios of periodogram values around the frequencies that are being tested as SSVEP stimuli (see \mydef{def:SNR} in \mysec{subsec:SNR}).
	
	\mysec{subsec:LM_etal} implements two versions of the well-known periodogram $F$-test frequency 
estimation method used in \cite{liavas-moustakides-1998,baka-tana-cich-neuro-let-2010}. The first version, which we call the \textbf{smoothed-\textit{F}} algorithm, is a direct implementation of \cite{liavas-moustakides-1998}. The second version, which we call the \textbf{GVZM-\textit{F}} algorithm, replaces a key data-estimated periodogram with the optimally-fitting GVZM PSD and making no other alterations. We compare the performance of the GVZM-\textit{F} and the smoothed-\textit{F} algorithms statistically.
	
	All four algorithms are used as SSVEP frequency estimators according to the protocol described in Appendix B. This Appendix also describes the procedures we used to generate the summary data statistics in \mytable{tab:SNR_vs_GVZM_summary} and \mytable{tab:LM_vs_GVZM_summary}.

\subsection{Set-up and Preprocessing}\label{subsec:Methods}

	As our data, we used the publicly-available  EEG recordings \cite{baka-tana-cich-neuro-let-2010} of four subjects undergoing a series of SSVEP experiments, using a 128-channel Biosemi active-electrode EEG system (\url{http://www.biosemi.com}) sampled at 256 Hz. Each subject experienced 15 25-second trials divided into five trials each of approximately 8 Hz, 16 Hz, and 28 Hz stimulation frequencies. Each 25-second trial consisted of a 5-second pre-stimulation epoch, a 15-second visual stimulation epoch, and a 5-second post-stimulation epoch. Further experimental conditions are presented in \cite{baka-tana-cich-neuro-let-2010}.
	
	We preprocessed each epoch separately. Using standard treatments for Biosemi recordings \cite{baka-tana-cich-neuro-let-2010}, quadratic trends in each channel were removed and the central  Cz channel (in 10/20-nomenclature was subtracted from all other electrodes. A virtual electrode over the visual cortex was created by averaging the Biosemi-nomenclature electrodes A14, A15, A16,  A21, A22, A23, A25, A27, A28, and A29. A virtual electrode close to the eye muscles was created by averaging the frontal 10/20 Fp2, Fpz, and Fp1 electrodes.The visual electrode was linearly regressed onto the eye electrode and the residual was used as the SSVEP response signal. This simple method of removing eyeblink artifacts worked well for our subjects.
	
	We excluded the mid-alpha (9.5 Hz--13.5 Hz) and high beta (23.5 Hz--26.5 Hz) non-stationary bands \cite{liavas-moustakides-1998} from all PSD functions, as well as frequencies below 6 Hz and above 50 Hz. This left 614 frequencies per periodogram for testing against the known stimuli.
	
	The cumulative distribution functions (CDFs) of the BCI-SNR statistic, required  in \mysec{subsec:SNR}, were calculated by bootstrap resampling \cite{efron-tibshirani-1993}. Specifically, for every one of the four subjects, each of their 15 pre-stimulation epochs was independently concatenated to their 15 post-stimulation epochs, yielding 225 sample baseline datasets which were then multiplied by a Tukey window with parameter 0.1 (using the window of \cite{kus-dusyk-etal-2013}). For each baseline, the BCI-SNR statistic was computed using equation Eq. \myeq{eq:SNR} at each of the 614 test frequencies resulting in 614 ``urns'', each urn containing the approximately 225 distinct SNR values which occured at that frequency. For each test frequency, 1000 samples (with replacement) of size 225 were then selected from its urn. Each of these 1000 samples generated its own CDF. Then these 1000 CDFs were averaged to obtain a representative CDF at that frequency. Repeating this procedure at every one of the 614 test frequencies yielded 614 empirical CDFs for each of the four subjects.
	
	The smoothed-\textit{F} algorithm \cite{liavas-moustakides-1998} of \mysec{subsec:LM_etal} estimated the expected PSD of the pre-stimulation epoch $x_\mathrm{pre}\Paren{n}$ by the smoothed periodogram approach of \cite{ljung-book-1987}. The circular autocorrelation \cite{oppen-schafer-book-1975} $R_\mathrm{pre}\Paren{m}$ of $x_\mathrm{pre}\Paren{n}$ was computed and then the DFT $S_\mathrm{pre}\Paren{k}$  of the windowed autocorrelation $h\Paren{m} \cdot R_\mathrm{pre}\Paren{m}$ was regarded as the expected PSD. The window $h\Paren{m}$ was a symmetric Hamming window of length $2M+1$, where $M$ is approximately 10\% of the data length of $x_\mathrm{pre}\Paren{n}$. In \cite{liavas-moustakides-1998}, the pre-stimulation data length was specifically chosen to be same as the stimulated epoch $x_\mathrm{stim}\Paren{n}$ so that their respective DFTs $S_\mathrm{pre}\Paren{k}$ and $S_\mathrm{stim}\Paren{k}$ could be compared easily at equal frequency indices $k$. Since our pre-stimulation epochs are shorter than the stimulation epochs, spline interperpolation of $S_\mathrm{pre}\Paren{k}$ was performed to resample it to the larger length. We found that the 10\% smoothed periodogram was sufficiently smooth that such resampling was very accurate.
	
	The authors of \cite{liavas-moustakides-1998} time-averaged multiple epochs to improve the SNR prior to detection. This required about 2 minutes of trial data, a very long duration for a practical real-time BCI.  For example, the longest epoch used by the well-known and successful SSVEP BCI of \cite{wangs-hong-gaos-2006} was 8 seconds, which was then continuously processed to yield average inter-selection intervals between 3.40 and 5.68 seconds. We therefore tested the GVZM-\textit{F} and smoothed-\textit{F} algorithms on the single-trial, unaveraged epochs of 15 seconds.
	
We performed the paired algorithm comparisons of GVZM-$\chi^2$ vs.  BCI-SNR in \mysec{subsec:SNR} and GVZM-\textit{F} vs. smoothed-\textit{F} in \mysec{subsec:LM_etal} by procedures outlined in Appendix B.  (See \myfig{fig:SNRTab_ROC}, \myfig{fig:LM_ROC}, \mytable{tab:SNR_vs_GVZM_summary}, and \mytable{tab:LM_vs_GVZM_summary}.)

	All fits of GVZM PSDs in \mydef{def:GVZM_atan} to actual EEG periodograms, used by both the GVZM-$\chi^2$ and GVZM-\textit{F} algorithms, were obtained by weighted least-squares optimization. We used weights proportional to $f^\beta$, for $\beta \approx 1.5$, where the $f$ are the frequencies over which we are optimizing. We observed that weighting increases the accuracy in the higher frequencies where the signal power is inherently small.
	
	Figures displaying spectra and the results of spectral tests show power density $S$ in dB relative to 1; that is 
$10 \, \mathrm{log}_{10}\Paren{S/1}$. However, all  actual critical values have been determined and hypothesis tests were performed in the original units of (amplitude unit)$^2$/Hz.


%

\subsection{Real-time Estimation of SSVEP Responses using the  GVZM-$\boldsymbol{\chi^2}$  Algorithm}\label{subsec:GVZM-Chi2}

	The most direct way to utilize the GVZM noise model in an estimation algorithm is by optimally fitting a GVZM PSD to recorded periodogram and calculating $\Paren{1/2} \, \chi^2\Paren{2}$ critical levels parallel to it. Recall that a GVZM-$\chi^2$-critical level at particular $P$-value is a curve parallel to $S_\mathrm{GVZM}(f)$,  showing the power below which periodogram values are confined with probability $1-P$, assuming the GVZM noise model to be correct. The frequencies of any spikes which extend above a pre-assigned $P$-value then are regarded as \myemph{positives}: frequencies at which the GVZM-$\chi^2$ algorithm will report the EEG as having true power, not merely random noise. All others are reported as negatives. In this way, stimulus frequency estimation is implemented by a sequence of hypothesis tests \cite{mood-graybill-boes-book-1974,benjamini-krieger-yekutieli-2006}, one at every frequency we intend to examine.
	
	\myfig{fig:chi2_8Hz} and \myfig{fig:chi2_28Hz} show the results of the GVZM-$\chi^2$ algorithm for SSVEP experiments at 8 Hz and 28 Hz respectively (subject 3, trial 2). In particular, according to the GVZM noise model, it is 99.5\% probable that a spectral spike will lie below the upper dashed critical level. These critical levels also display how accurately the GVZM noise model fits the spectra of recorded EEG background processes.
	
	It is clear from \myfig{fig:chi2_8Hz} that the GVZM-$\chi^2$ algorithm accurately estimated the 8Hz fundamental SSVEP response and its two harmonics at significance level $P = 0.005$, thus 
generating no false negatives (also called) Type II errors\cite{mood-graybill-boes-book-1974}). Moreover, it has correctly excluded all other spikes as simply random noise and thus avoided all false positives. (Also called Type I errors \cite{mood-graybill-boes-book-1974} or false discoveries \cite{benjamini-krieger-yekutieli-2006}).
	
	In \myfig{fig:chi2_28Hz} the 28 Hz response was estimated accurately however there are several false positives
arising, because of non-stationary power in the alpha- and beta-bands (approximately 25 Hz). As discussed in \mysec{subsec:Methods}, such bands need to be pre-excluded from the frequencies to be tested.

\subsection{SSVEP Frequency Estimation Using the GVZM-$\chi^2$ and BCI-SNR Algorithms}\label{subsec:SNR}

	In this section, we describe the BCI-SNR algorithm and compare it to the GVZM-$\chi^2$ algorithm.
	
	The BCI-SNR statistic for SSVEP procedures was first defined in \cite{wangs-hong-gaos-2006}, where it was used as a simple measure of signal strength for determining optimal stimulus frequencies. It subsequently became a popular frequency estimator for SSVEP BCIs \cite{kus-dusyk-etal-2013} and more general applications \cite{vialatte-maurice-dauwels-cichocki-2010}.  The phrase ``power spectral density analysis'' (PSDA) also has been used \cite{wei-xiao-lu-2011} for methods based on the BCI-SNR. 

\mybegdef{def:SNR}
	The BCI-SNR statistic \cite{wangs-hong-gaos-2006} of a signal $x$ at the $k^\mathrm{th}$ test frequency $f_k$ is the ratio
\mybegeq{eq:SNR}
	SNR_x\Paren{f_k} = \frac{n \cdot \hat{S}_x\Paren{f_k}}{\sum\limits_{\substack{j = -n/2 \\ j\ne0}}^{n/2}
			{  \hat{S}_x\Paren{f_k+j\cdot\Delta f} } },
\myendeq

\noindent where $\hat{S}_x$ is an estimator of the sample spectrum, $\Delta f$ is the spectral resolution of the estimated frequency domain, and $n$ is a small integer. To be consistent with \cite{kus-dusyk-etal-2013}, we use $n = 6$.
\myenddef

	The BCI-SNR statistic is sometimes used as a non-blind detector for a short list $f_1,\ldots, f_K$ of known  SSVEP BCI selection frequencies. The subject's selection is considered to be that frequency $f_k$ which has the largest $SNR_x$ value; i.e., 
$f_{\mathrm{selected}} = \mathrm{argmax}_{f_k} \; SNR_x\Paren{f_k}$. (For example, \cite{wei-xiao-lu-2011,kus-dusyk-etal-2013}.)
	
	To use the BCI-SNR algorithm as a blind SSVEP frequency estimator, as described by Appendix B, we require each individual probability distribution function of $SNR_x\Paren{f_k}$ for every test frequency $f_k$. These were estimated by bootstrap resampling as described in \mysec{subsec:Methods}.


\begin{figure} [h!]
\centering
\vspace{-.01\textwidth}
\includegraphics[width=0.43\textwidth,left]{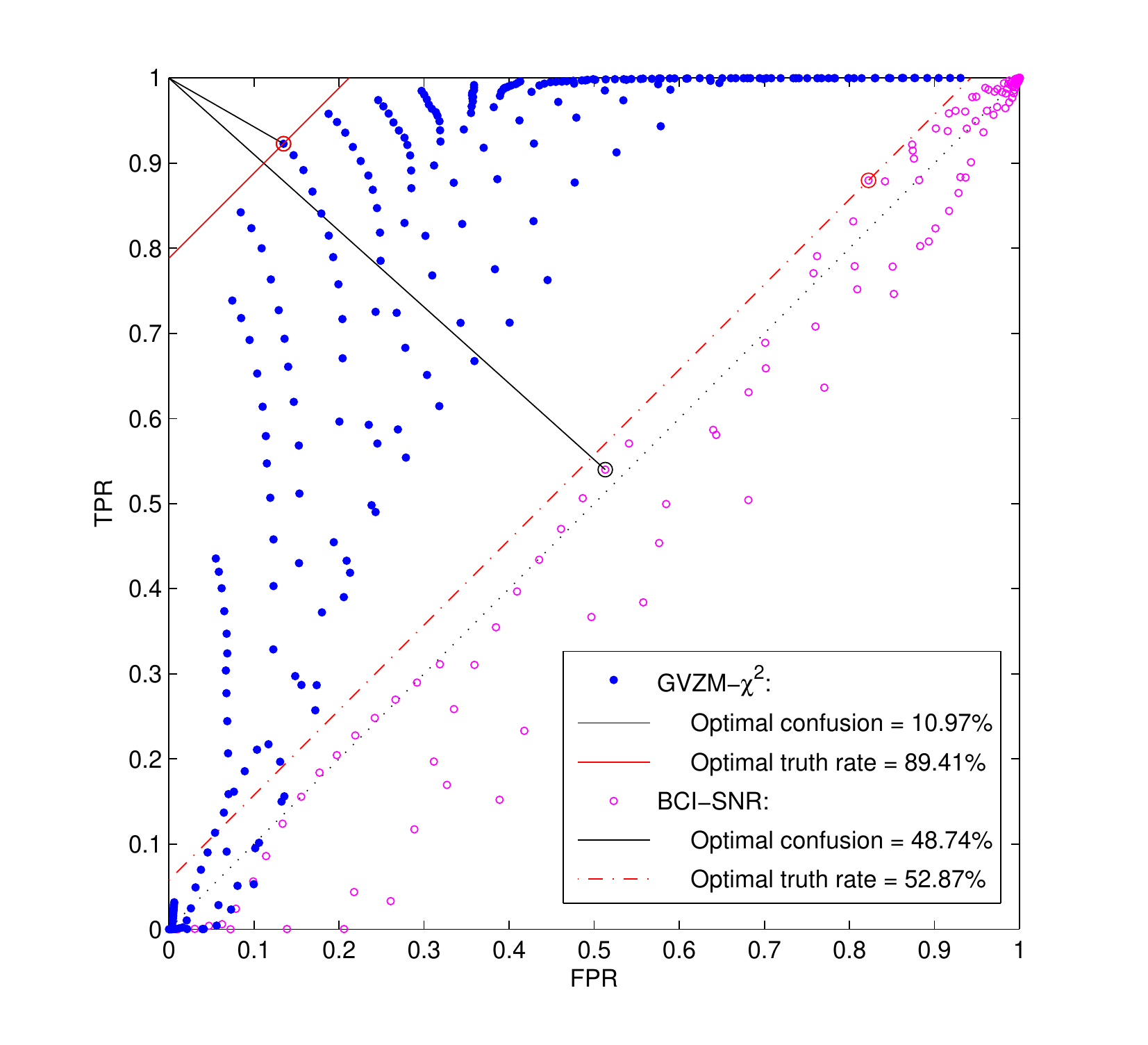}
\vspace{-.05\textwidth}
\caption{GVZM-$\chi^2$ vs. BCI-SNR example: the minimum-variance, unbiased, single-trial ROC estimator, showing optimal operating points, using the 28 Hz data of Fig. 1.}
\label{fig:SNRTab_ROC}
\vspace{-.04\textwidth}
\end{figure}

	\myfig{fig:SNRTab_ROC} shows an example of a minimum-variance, unbiased, single-trial estimate of the comparison \myemph{receiver operating characteristics (ROC)} graph \cite{fawcett-2006} as described in Appendix B. The dataset is that of \myfig{fig:RawEEG28Hz} (subject 3, trial 1, 28 Hz stimulus). 
	
	In \myfig{fig:SNRTab_ROC}, TPR and FPR denote the true and false positive rates, calculated at 256 operating points as detailed in the Appendix. Confusion measures the fractional distance of an operating point from ideal performance TPR = 1, FPR = 0 while the truth rate is a weighted average of the TPR and the true negative rate TNR = (1 - FPR) intended as a measure of accuracy
\[\left\{
\begin{aligned}
	 \mathrm{Confusion} &= \sqrt{\Paren{1-\mathrm{TPR}}^2 + \mathrm{FPR}^2} / \sqrt 2 \\
		\mathrm{Truth \; Rate} &= \Paren{1-p_0}\cdot\mathrm{TPR} + p_0\cdot\Paren{1-\mathrm{FPR}},
\end{aligned}
\right.\]
where $0 \myle p_0 \myle 1$. In a Bayesian situation, with $p_0$ the probability of the null hypothesis \cite{mood-graybill-boes-book-1974}, the truth rate is the probability of a true decision. We assume $p_0 = 1/2$.
	
	We observe that the GVZM-$\chi^2$ algorithm outperforms the BCI-SNR algorithm at nearly every operating point. Moreover, even on this very difficult 28Hz stimulus, the optimal operating point for GVZM-$\chi^2$ identifies the true stimulus frequency with probability above 90\% with FPR below 20\%.

\begin{table}[!h]
\vspace{-.02\textwidth}
\caption{Performance improvement of GVZM-$\chi^2$ over BCI-SNR}
\label{tab:SNR_vs_GVZM_summary}
\centering
\resizebox{.49\textwidth}{!}{
\begin{tabular}{ | c | c | c | c | c | c | c |}
\hline
\parbox[c][2.5em][c]{.11\textwidth}{\textit{Combined Optimal Results}} & \textbf{\%} & \parbox[c][2.5em][c]{.07\textwidth}{\textbf{Unconfused\\Trials}} 
	& \textbf{Pooled SE} & \textbf{\textit{t}-score} & \textbf{df} & \textbf{\textit{P}-value} \\ \hline
\parbox[c][2.0em][c]{.12\textwidth}{\textbf{Confusion Decrease}} & 29.77 & 34 & 0.0371 & 3.253 & 11 & 0.004 \\ \hline
\parbox[c][2.0em][c]{.12\textwidth}{\textbf{TruthRate Increase}} & 17.92 & 34 & 0.0356 & 3.133 & 11 & 0.005 \\ \hline
\end{tabular}
} 
\end{table}

	\mytable{tab:SNR_vs_GVZM_summary} summarizes the pooled ROC results for $N = 34$ trials, in which at least one of the algorithms had confusion below 35\% (``unconfused'' trials by definition). Pooled SE denotes the standard error
\[
	\mathrm{Pooled \; SE} =\sqrt{\Paren{\sigma_\mathrm{GVZM}^2 + \sigma_\mathrm{SNR}^2} / N}
\]
appropriate to the \textit{t}-test for the difference of means.
	
	\mytable{tab:SNR_vs_GVZM_summary} shows that  the GVZM-$\chi^2$ algorithm outperforms the BCI-SNR algorithm on both the confusion and truth rate measures with statistical significance above 99\%.

\subsection{SSVEP Frequency Estimation Using the GVZM-\textit{F} and Smoothed-\textit{F} Algorithms}\label{subsec:LM_etal}

	In \cite{liavas-moustakides-1998}, the authors assume, based on the theory of periodograms developed in detail in \cite{brockwell-davis-book-1991}, that the periodogram random process $S_x\Paren{k}$, $0 \myle k \myle N-1$ of the EEG background time series $x\Paren{n}$ is given asymptotically by 
\mybegeq{eq:Liavas_etal}
	S_x\Paren{k} = \Expv\Brack{S_x\Paren{k}} \cdot \Paren{1/2} \, \Xi \Paren{k},
\myendeq
where $\Xi\Paren{k}$, $0 \myle k \myle N-1$ is a process with $\Xi\Paren{k} \sim \chi^2\Paren{2}$ and which are independent for $0 \myle k, l < N/2$ with $k \ne l$. 

	If the function $\Expv\Brack{S_x\Paren{k}}$ were known, then \myeq{eq:Liavas_etal} implies that the test statistics
\[
	s_x\Paren{k} \Eqdef 2 \cdot S_x\Paren{k} /  \,\Expv\Brack{S_x\Paren{k}}
\]
for $0 \myle k \myle N-1$ would be iid $\chi^2\Paren{2}$ random variables.

	A fixed set of indices $\Omega \subseteq \Curly{0,1,\cdots,N-1}$ is selected to represent what we regard as stationary frequencies; e.g., alpha-band frequencies are excluded (cf. \mysec{subsec:Methods}). Then, under the null hypothesis that there is no SSVEP spike at frequency index  $k_\mathrm{test}$ we have
\begin{align}\label{eq:F-stat}
	&\frac {\sum_{k \in \Omega_\mathrm{test}} {\! s_x\Paren{k}} \;/\; N_\mathrm{test}}
		{ \sum_{k' \in \Omega \setminus \Omega_\mathrm{test}} {\! s_x\Paren{k'}} \;/ \Paren{N_\Omega - N_\mathrm{test}}}  \\
	&\qquadFour \sim F\Paren{N_\mathrm{test}, N_\Omega - N_\mathrm{test}}\nonumber,
\end{align}
where $\Omega_\mathrm{test}$ is the set of indices $k_\mathrm{test}$ and its harmonics in $\Omega$, $N_\mathrm{test}$ is the size of $\Omega_\mathrm{test}$, $N_\Omega$ the size of $\Omega$, and $F\Paren{d_1,d_2}$ is the $F$--distribution with degrees of freedom $d_1, \, d_2$ \cite{mood-graybill-boes-book-1974}.

	The key issue is how to obtain the function $\Expv\Brack{S_x\Paren{k}}$. In \cite{liavas-moustakides-1998}, the authors estimated this function by computing a ``smoothed periodogram'' $S_\textrm{smooth}\Paren{k}$ of a pre-stimulation epoch as described in \mysec{subsec:Methods} and used $S_\textrm{smooth}\Paren{k}$ as a substitute for $\Expv\Brack{S_x\Paren{k}}$. This constituted their smoothed-\textit{F} algorithm.
	
\begin{figure} [h!]
\centering
\vspace{-.02\textwidth}
\includegraphics[width=0.45\textwidth]{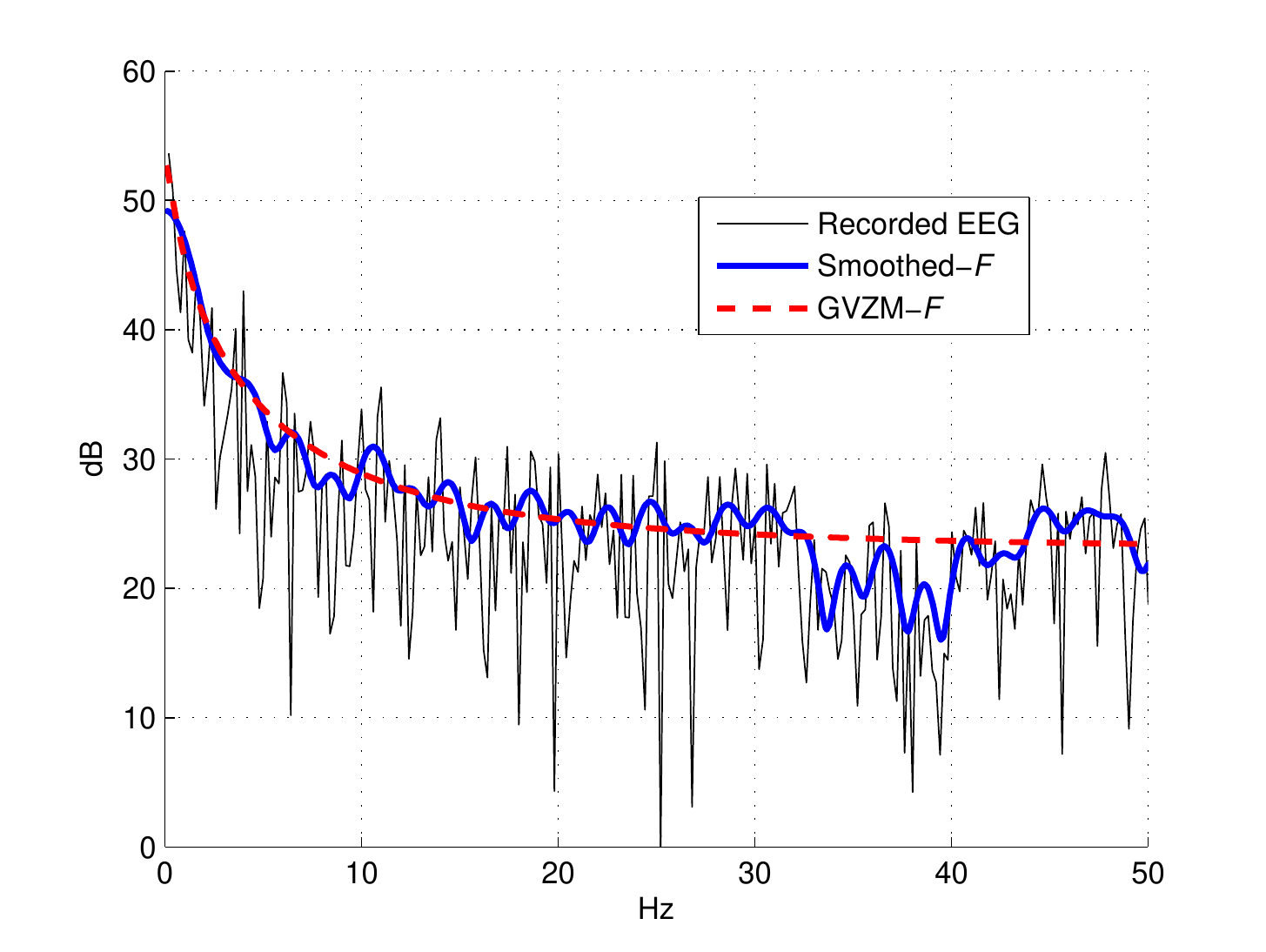}
\vspace{-.02\textwidth}
\caption{Baseline (non-stimulus) PSDs used as the 28 Hz, $\chi^2$ references for the smoothed-\textit{F} and GVZM-\textit{F} algorithms.}
\label{fig:baseline_28Hz}
\vspace{-.02\textwidth}
\end{figure}

	However, by using the GVZM noise model, we expect to obtain a more accurate baseline estimate by using the GVZM PSD which optimally fits the same pre-stimulus epoch as a substitute for $\Expv\Brack{S_x\Paren{k}}$. This is our GZVM-\textit{F} algorithm. \myfig{fig:baseline_28Hz} shows the periodogram from the pre-stimulus epoch of the data of \myfig{fig:RawEEG28Hz} (subject 3, trial 1, 28Hz stimulus) with both the smoothed periodogram and the fitted GVZM PSD displayed. The two algorithms are compared as described in Appendix B. \myfig{fig:LM_ROC} shows an example of a minimum-variance, unbiased, single-trial estimate of the comparison ROC graph.
	

	The dataset is that of \myfig{fig:RawEEG28Hz} (subject 3, trial 1, 28 Hz stimulus). As in \mysec{subsec:SNR}, the GVZM-\textit{F} algorithm outperforms the smoothed-\textit{F} algorithm at nearly every operating point, achieving an optimal performance of over 95\% probability of true identification with just above 20\% FPR.

\begin{table}[!h]
\vspace{-.01\textwidth}
\caption{Performance improvement of GVZM-\textit{F} over Smoothed-\textit{F}}
\label{tab:LM_vs_GVZM_summary}
\centering
\resizebox{.49\textwidth}{!}{
\begin{tabular}{ | c | c | c | c | c | c | c |}
\hline
\parbox[c][2.5em][c]{.11\textwidth}{\textit{Combined Optimal Results}}  & \textbf{\%} & \parbox[c][2.5em][c]{.07\textwidth}{\textbf{Unconfused\\Trials}} 
	& \textbf{Pooled SE} & \textbf{\textit{t}-score} & \textbf{df} & \textbf{\textit{P}-value} \\ \hline
\parbox[c][2.0em][c]{.12\textwidth}{\textbf{Confusion Decrease}} & 30.57 & 56 & 0.0322 & 2.901 & 11 & 0.007 \\ \hline
\parbox[c][2.0em][c]{.12\textwidth}{\textbf{TruthRate Increase}} & 12.67 & 56 & 0.0278 & 3.234 & 11 & 0.004 \\ \hline
\end{tabular}
} 
\vspace{-.02\textwidth}
\end{table}

	\mytable{tab:LM_vs_GVZM_summary} summarizes the pooled ROC results as described in Appendix B for the 56 trials in which at least one of the algorithms had confusion below 35\%. The large number of unconfused trials of these algorithms is a result of the inherent stability of the carefully-designed underlying statistic \myeq{eq:F-stat}.
	
	\mytable{tab:LM_vs_GVZM_summary} shows that  GVZM-\textit{F} algorithm outperforms the smoothed-\textit{F} algorithm on both the confusion and accuracy measures with statistical significance above 99\%. 
	

\begin{figure} [h!]
\vspace{-.02\textwidth}
\includegraphics[width=0.43\textwidth,left]{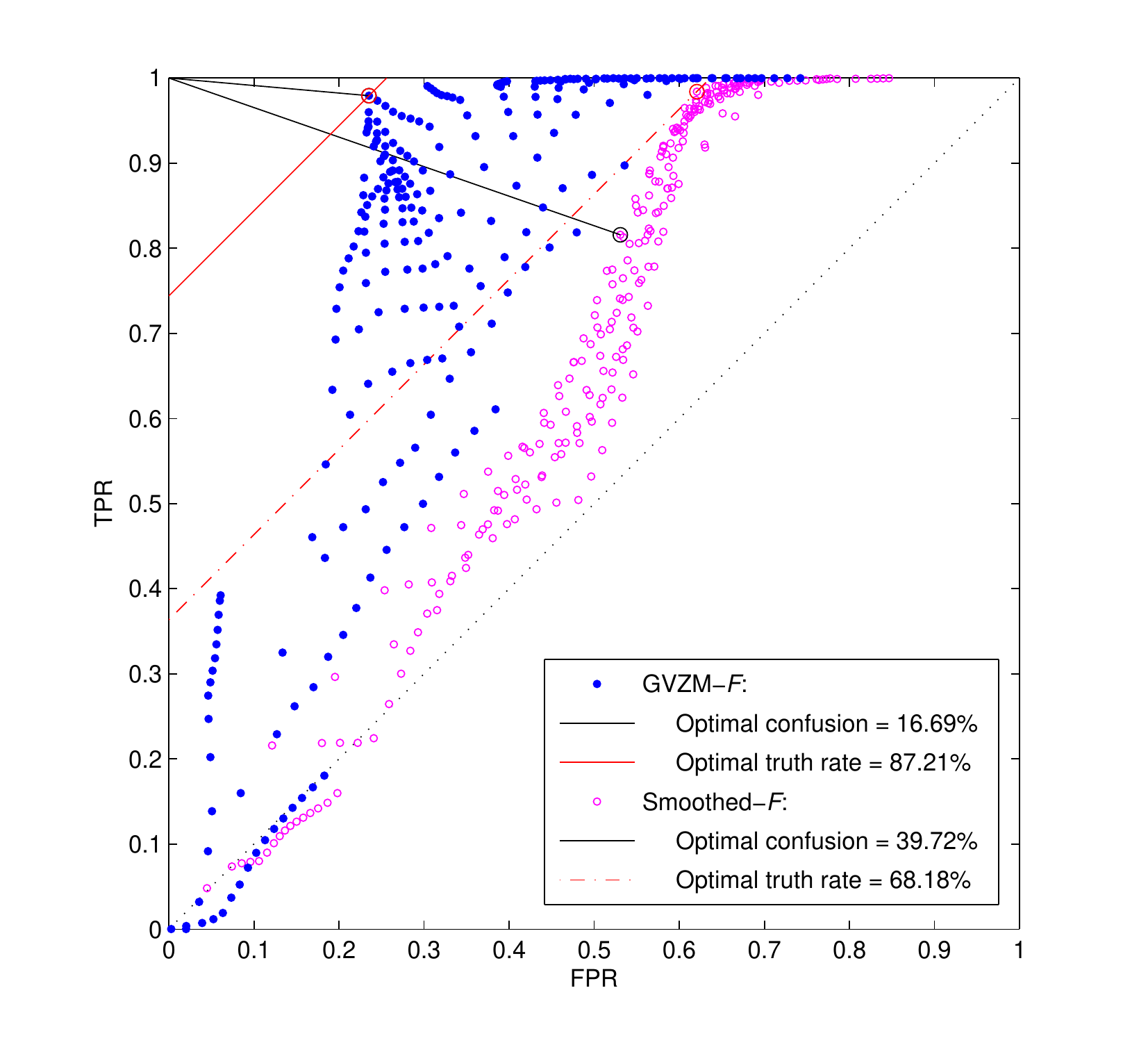}
\vspace{-.05\textwidth}
\caption{GVZM-\textit{F} vs. Smoothed-\textit{F} example: the minimum-variance, unbiased, single-trial ROC estimator, showing optimal operating points, using the 28 Hz data of Fig. 1.}
\label{fig:LM_ROC}
\vspace{-.04\textwidth}
\end{figure}


\section{Conclusions}\label{sec:Summary}

	In this paper, we showed the necessity of accurate statistical models of EEG background noise for applications to neuroscience, neurology, and real-time BCIs.
	
	Based on our analysis and numerical experiments, we proposed a specific, five-parameter statistical family of EEG background periodograms using the GVZM PSD function.  To the best of our knowledge, the GVZM PSD function is the only EEG noise PSD model with a fixed number of parameters, matches recorded EEG PSD's with high accuracy over the full spectrum from 0 Hz to 30 Hz, and has approximately $1/f^\theta$ behavior in the mid-frequencies without infinities.
	
	We validated this model using three complementary approaches. First, we briefly discussed our theoretical work \cite{paris-atia-vosoughi-berman-2015a} on neurological ion channels and quantum stochastic processes and noted that this theory implies the applicability of the GVZM PSD function to large networks of ion channels in maximum entropy statistical equilibrium. Second, we presented  a class of time series we call  AR-GVZM, processes which simulate EEG background noise. We showed that the periodograms of AR-GVZM, processes approach the GVZM PSD function as the number of subprocesses becomes large. 

Third, we designed two algorithms to estimate SSVEP frequencies for real-time BCI applications and applied them to a public set of SSVEP recorded data. We showed statistically that both GVZM-based algorithms were more accurate than two commonly used alternatives, even on difficult data.
	
	We conclude that the GVZM noise model accurately characterizes the statistics of EEG background periodograms and, therefore, is a very suitable model for algorithm design, brain signal simulation, and neuroscience investigations. In addition, we believe our applied results indirectly validate quantum ion channel maximum entropy analysis, which is the theoretical foundation of the GVZM noise model.

\section{Appendices}\label{sec:Appendices}

Appendix A presents the background for a useful class of random processes we call $ARMA(P,Q,N)$ processes which provide motivation for the GVZM model of \mysec{sec:GVZM}. Appendix B is an overview of the statistical methodology used for the algorithm comparisons of \mysec{sec:SSVEP}.


\begin{figure*}
	\vspace{-.03\textwidth}
	\centering
	\includegraphics[width=.68\textwidth]{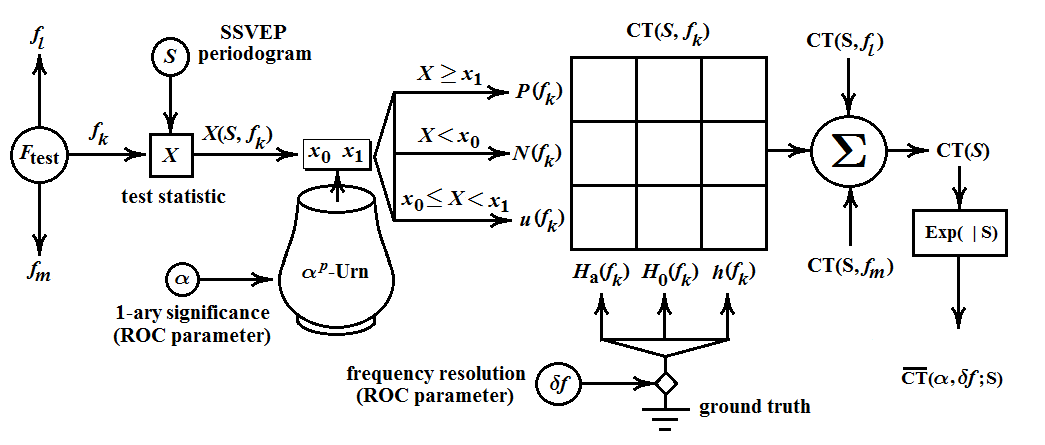}
	\vspace{-.01\textwidth}
	\caption{Flow chart for the calculation of $\overline{\mathrm{CT}}\Paren{\alpha,\Delta f ; S}$: the minimum variance, unbiased, single-trial estimator of the $3 \times 3$ contingency table of the input periodogram $S$, with ROC control parameters $\alpha$, $\Delta f$.}
	\label{fig:CTchart}
	\vspace{-.04\textwidth}
\end{figure*}

\subsection{The Periodogram Distribution of Periodic ARMA Processes}\label{sec:ARMA_N}

\mybegdef{def:N_ARMA} A periodic Gaussian $\boldsymbol{ARMA\Paren{P,Q,N}}$ \myemph{process} is a zero-mean, stationary random process $\Curly{x\Paren{n}}_{n = -\infty}^\infty$ satisfying
\mybegeq{eq:ARMA_N}
	\sum_{p = 0}^P{a_p \, x\Paren{n-p}} = \sum_{q = 0}^Q{b_q \, \nu\Paren{n-q}},
\myendeq
\noindent where $a_0 = b_0 = 1$ and  $\Curly{\nu\Paren{n}}_{n = -\infty}^\infty$ is an iid, zero-mean, Gaussian process which is $N$-periodic: $\nu\Paren{n+N} = \nu\Paren{n}$, for all $n$, and the polynomials $A\Paren{z} = \sum_{p=0}^P{a_p z^p}$ and $B\Paren{z} = \sum_{q=0}^Q{b_q z^q}$ have no common zeros.
The equality in the periodicity condition is meant to be exact; i.e., there are really only $N$ distinct random processes $\nu\Paren{0}, \ldots, \nu\Paren{N-1}$.
\myenddef 

\mybegdef{} An $ARMA\Paren{P,Q,N}$ system is \myemph{causal} \cite{brockwell-davis-book-1991} if the polynomial  $A\Paren{z}$ defined above has no zeros on or inside the unit circle.
\myenddef

\mybegthm{thm:ChiTwo} Let $x\Paren{0}, \ldots, x\Paren{N-1}$ be one period of a causal Gaussian $ARMA\Paren{P,Q,N}$ process and let $S_x\Paren{k}$, $0 \myle k \myle N-1$, be its periodogram (\mydef{def:periodogram}). Then
\[
	S_x\Paren{k} = 2\pi \sigma_\nu^2 \cdot \frac{\Abs{B\Paren{\Exp{-i\Paren{2\pi k/N}}}}^2}{\Abs{A\Paren{\Exp{-i\Paren{2\pi k/N}}}}^2} \cdot \Paren{1/2} \, \Xi\Paren{k},
\]
\noindent where $\Xi\Paren{k}$,  $0 \myle k \myle N-1$, is  a sequence of $\chi^2\Paren{2}$ distributed \cite{mood-graybill-boes-book-1974} random variables which are independent for $0 \myle k, l < N/2$ when $k \ne l$ and $\sigma_\nu^2$ is the variance of $\nu\Paren{n}$ .
\myendthm

	Because of space limitations, we omit the proof of this result, which is a simple application of the techniques of \cite{brockwell-davis-book-1991}.

\subsection{Comparison Protocols for SSVEP Estimation Algorithms}\label{subsec:Protocols}

	To simplify analysis of the algorithms of \mysec{sec:SSVEP}, all four were reduced to a common $m$-ary test \cite{benjamini-krieger-yekutieli-2006} protocol, outlined by the flow chart \myfig{fig:CTchart}.
	
	Every $f \in F_\mathrm{test}$ in a given set of \myemph{test frequencies} is to be independently judged by the algorithm for its presence or absence as an SSVEP stimulus in the experimental periodogram $S$. No constraint is placed on the number of test frequencies $f \in F_\mathrm{test}$ the algorithm may report as present (positives, P), absent (negatives, N), or undetermined (u). Both pairs in a comparison (GVZM-$\chi^2$ vs. BCI-SNR and GVZM-$F$ vs. smoothed-$F$) are tested against the same \myemph{ground truth}, which has access to the true SSVEP stimulus frequencies and possible harmonics for this experiment. Each algorithm is characterized by a \myemph{test statistic} $X\Paren{\boldsymbol{\mathrm{S}},f}$ with CDFs 
\[
	P_X\Paren{x,f} \Eqdef \Prob\Brack{X\Paren{\boldsymbol{\mathrm{S}},f} \myle x},
\]
where $\boldsymbol{\mathrm{S}}$ is a random variable varying over all possible periodograms that $X\Paren{\Hole,f}$ could receive ($S$ is a particular value of $\boldsymbol{\mathrm{S}}$). We assume that the larger the value of $X\Paren{\boldsymbol{\mathrm{S}},f}$, the more likely $f$ is a stimulus frequency.

\textbf{GVZM-$\boldsymbol{\chi^2}$}: $X\Paren{\boldsymbol{\mathrm{S}},f} = \boldsymbol{\mathrm{S}}\Paren{f}/S_\mathrm{GVZM}\Paren{f}$ and $P_X\Paren{x,f}$ is given by the inverse $\Paren{1/2}\cdot\chi^2\Paren{2}$ distribution (see Eq. \myeq{eq:GVZM_process}).

\textbf{BCI-SNR}: $X\Paren{\boldsymbol{\mathrm{S}},f}$ is given by Eq. \myeq{eq:SNR} and the CDFs are obtained empirically by bootstrap resampling as described in \mysec{subsec:Methods}.

\textbf{GVZM-\textit{F}}: $X\Paren{\boldsymbol{\mathrm{S}},f}$ is the ratio Eq. \myeq{eq:F-stat}, with $\Expv\Brack{S_x\Paren{k}}$ the GVZM fit to the baseline data. The CDFs are the inverse \textit{F} distribution with appropriate degrees of freedom.

\textbf{smoothed-\textit{F}}: $X\Paren{\boldsymbol{\mathrm{S}},f}$ is the ratio Eq. \myeq{eq:F-stat}, with $\Expv\Brack{S_x\Paren{k}}$ the smoothed periodogram of \cite{ljung-book-1987}.The CDFs are the inverse \textit{F} distribution with appropriate degrees of freedom.

	Let $0 \myle \alpha \myle 1$ be the \myemph{1-ary significance level}. The $\boldsymbol{\alpha^p}$\myemph{-urn} for $X\Paren{\boldsymbol{\mathrm{S}},f}$ is a conceptual container of paired values $x_0 \myle x_1$ whose likelihood of occurrence is determined by $P_X\Paren{x,f}$. The $\boldsymbol{\alpha^p}$\myemph{-urn test} is the following randomized procedure: Calculate $X\Paren{\boldsymbol{\mathrm{S}},f}$ and independently select a pair $x_0, x_1$ from the $\alpha^p$-urn. Then $f$ is adjudged present in $\boldsymbol{\mathrm{S}}$ if $X\Paren{\boldsymbol{\mathrm{S}},f} \myge x_1$, absent if  $X\Paren{\boldsymbol{\mathrm{S}},f} < x_0$, and undetermined otherwise.

	Ground truth answers questions about binary hypotheses $H_0\Paren{f},H_\mathrm{a}\Paren{f}$, where the null hypothsis $H_0\Paren{f}$ is: frequency $f$ is not an SSVEP stimulus. It may fail to answer, which we symbololize in \myfig{fig:CTchart} by ``$h\Paren{f}$''. Since no estimation algorithm can be 100\% accurate, ground truth must contain a parameter $\delta f$ which determines how much error $\Abs{f - f_\mathrm{ssvep}}$ it will tolerate between a test frequency $f$ and a known stimulus or harmonic frequency $ f_\mathrm{ssvep}$. We have found SSVEP estimation to be very sensitive to $\delta f$ and its optimal value must be determined for each subject during pre-test training.
	 
	 With these definitions, the \myemph{single-trial contingency table} $CT\Paren{\boldsymbol{\mathrm{S}}}$ can be calculated as shown in \myfig{fig:CTchart}.  Note that multiple true positives are possible, even with a single stimulus frequency $f_\mathrm{ssvep}$, both because ground truth may report harmonics of $f_\mathrm{ssvep}$ and because the resolution $\delta f > 0$ must report test frequencies in an interval around $f_\mathrm{ssvep}$. When a particular periodogram $\boldsymbol{\mathrm{S}} = S$ is analyzed, the conditional expectation $\overline{CT}\Paren{S} \Eqdef \Expv\Brack{CT\Paren{\Hole} \st S}$, which is known to be the minimum variance unbiased estimator of $CT\Paren{\Hole}$ \cite{mood-graybill-boes-book-1974}, can be calculated.  From the cells of $\overline{CT}\Paren{S}$ , we can derive the TPR and FPR (\mysec{subsec:SNR}) as $\mathrm{TPR} \Eqdef \mathrm{TP} / N_\mathrm{a}$, $\mathrm{FPR} \Eqdef \mathrm{FP} / N_0$, where TP, FP are the number of true and false positives and $N_\mathrm{a}$,$N_0$ are the number of true hypotheses $H_\mathrm{a}$,$H_0$ respectively.
	 
	 As seen in \myfig{fig:CTchart}, the estimator $\overline{CT}\Paren{\alpha,\delta f; S}$ depends on the parameters $\alpha,\delta f$. By varying these, we can plot the resulting operating points $\Paren{\mathrm{FPR}\Paren{\alpha,\delta f},\mathrm{TPR}\Paren{\alpha,\delta f}}$ on a \myemph{single-trial ROC} graph \cite{fawcett-2006}. The ROC determines the optimal values of $\alpha,\delta f$ for the periodogram $S$. \myfig{fig:SNRTab_ROC} and \myfig{fig:LM_ROC} show the operating points of 256 choices of $\alpha,\delta f$ for the various algorithms, on the \myfig{fig:RawEEG28Hz} periodogram, as well as the optimal operating points according to the two accuracy measures described in \mysec{subsec:SNR}. 
	 
	 By creating single-trial ROCs for all periodograms $S$ in an experiment, the statistical distribution of optimal accuracy measures can be inferred. \mytable{tab:SNR_vs_GVZM_summary} and \mytable{tab:LM_vs_GVZM_summary} show the analysis of $N = 12$ optimal operating point measures corresponding to the three subjects and four possible stimulus frequencies, each of which were tested in five trials (\mysec{subsec:Methods}).

\section*{References}
\bibliographystyle{IEEETran}
\bibliography{\BibsDir/TMBE.arXiv} 

\end{document}